\documentclass[aps, pra, twocolumn, 10point, longbibliography,  citeautoscript]{revtex4-1}
\usepackage{amsmath}
\usepackage{amssymb}
\usepackage{subfigure}
\usepackage{color}
\usepackage{blkarray}
\usepackage{dsfont}
\usepackage{mathrsfs}
\usepackage[breaklinks=true,colorlinks=true,linkcolor=blue,urlcolor=blue,citecolor=blue]{hyperref}
\usepackage{bbm}
\usepackage{hyperref}
\usepackage{amsthm}
\usepackage{graphicx, nicefrac}
\usepackage{enumitem}
\usepackage{xspace}
\usepackage{comment}
\usepackage[dvipsnames]{xcolor}
\usepackage{txfonts}

\def\bra#1{{\left\langle #1 \right|}}
\def\ket#1{{\left| #1 \right\rangle}}

\newcommand{\Id}{\mathds{1}}

\newcommand{\cnot}{\ensuremath{\mathsf{CNOT}}\xspace}
\newcommand{\cphase}{\ensuremath{\mathsf{CPHASE}}\xspace}
\renewcommand{\vec}[1]{\mathbf{#1}}

\newcommand{\ie}{i.e.}
\newcommand{\eg}{e.g.}

\begin{document}
\title{Measuring the Capabilities of Quantum Computers}
\author{Timothy Proctor\textsuperscript{*}} 
\author{Kenneth Rudinger}
\author{Kevin Young}
\author{Erik Nielsen}
\author{Robin Blume-Kohout}
\affiliation{Quantum Performance Laboratory, Sandia National Laboratories, Albuquerque, NM 87185, USA and Livermore, CA 94550, USA}
\date{\today} 
\begin{abstract}
{ 
A quantum computer has now solved a specialized problem believed to be intractable for supercomputers, suggesting that quantum processors may soon outperform supercomputers on scientifically important problems. But flaws in each quantum processor limit its capability by causing errors in quantum programs, and it is currently difficult to predict what programs a particular processor can successfully run. We introduce techniques that can efficiently test the capabilities of any programmable quantum computer, and we apply them to twelve processors. Our experiments show that current hardware suffers complex errors that cause structured programs to fail up to an order of magnitude earlier --- as measured by program size --- than disordered ones. As a result, standard error metrics inferred from random disordered program behavior do not accurately predict performance of useful programs. Our methods provide efficient, reliable, and scalable benchmarks that can be targeted to predict quantum computer performance on real-world problems.
}
\end{abstract}
\maketitle
\noindent
Quantum processors are on the verge of realizing their promise to revolutionize computing. A quantum processor has now executed programs believed to defy classical simulation \cite{arute2019quantum}, and many hybrid quantum/classical algorithms have appeared that offer the possibility of near-term computational advantage \cite{preskill2018quantum}. Publicly available quantum processors continue to proliferate, and with them a widespread interest in running application-inspired quantum programs. But contemporary quantum processors are plagued by errors that will cause many of these programs to fail. Existing tools for characterization and benchmarking \cite{arute2019quantum, neill2018blueprint, boixo2018characterizing, cross2018validating, magesan2011scalable, emerson2007symmetrized, emerson2005scalable, proctor2018direct, linke2017experimental, harrow2017quantum, wright2019benchmarking, erhard2019characterizing,  flammia2019efficient} probe the magnitude and type of these errors. But none of them provide direct insight into a processor's capability --- the programs it can run successfully --- and most are not practical on devices that are large enough to potentially demonstrate a quantum advantage. In this work we introduce the first scalable benchmark that is able to efficiently probe and summarize the capability of any gate-model quantum computer, and we present the first systematic study of the capabilities of publicly accessible quantum processors.

The errors suffered by multi-qubit quantum processors are complex and varied, often including effects such as crosstalk \cite{gambetta2012characterization}, coherent noise \cite{huang2019performance, kueng2015comparing, murphy2019controlling}, and drift \cite{mavadia2017experimental, proctor2019detecting}. Simple models for device performance that ignore this complexity offer inaccurate predictions, while complex models are generally intractable to learn or computationally taxing to use. Instead, we argue that the capability of a processor is best probed by running a set of representative test quantum programs whose measured output can be verified classically. 

\begin{figure*}[t!]
\includegraphics[width=18cm]{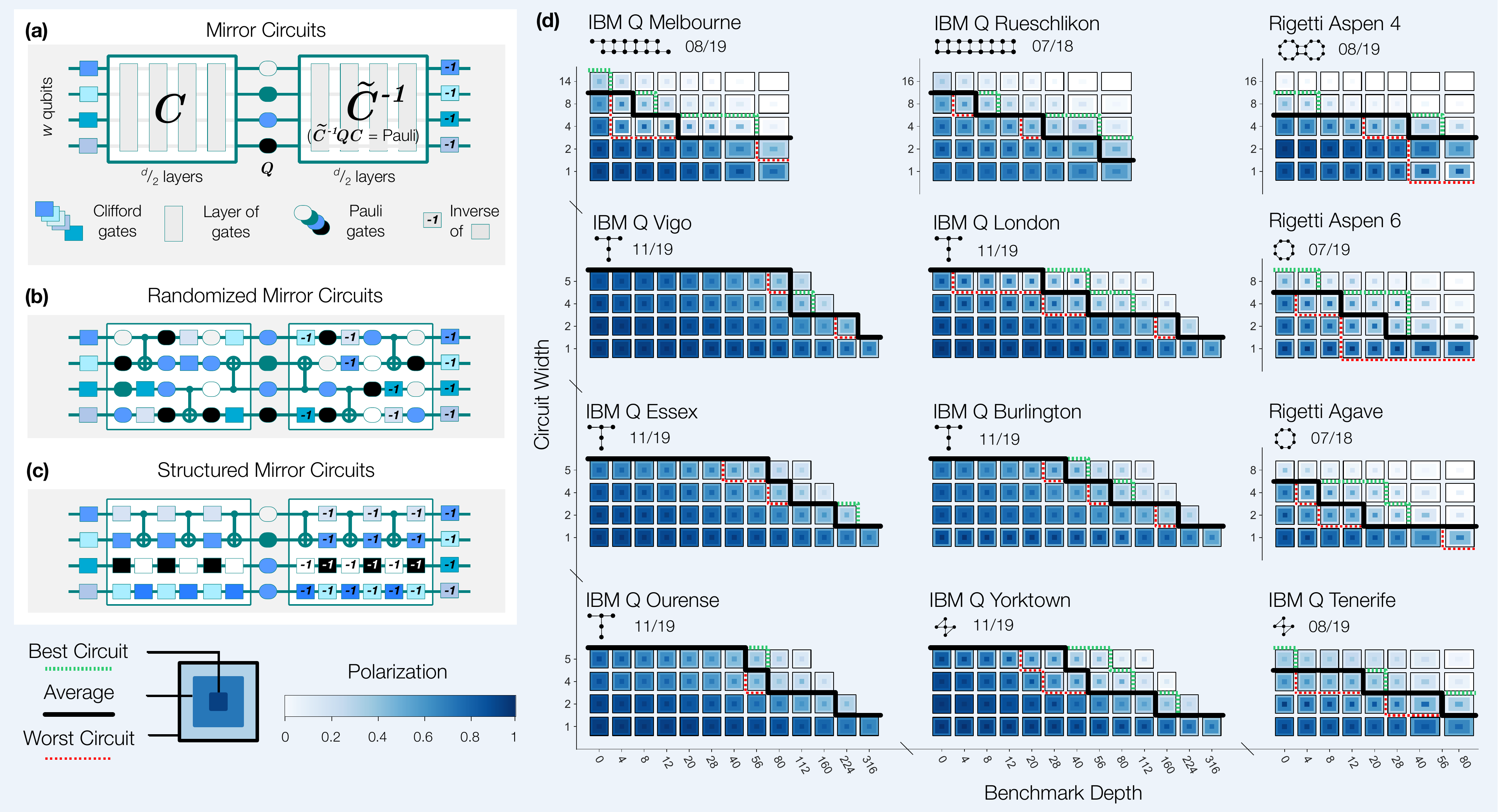}
\caption{{\bf A scalable method for benchmarking a quantum computer's capability.}~{\bf (a)} Mirror circuits --- quantum circuits with a reflection structure --- can be used to efficiently benchmark arbitrarily large quantum computers, because without errors they output a unique and easy-to-calculate bit string. Mirror circuits can be constructed from {\bf (b)} random disordered logic gate sequences; {\bf (c)} ordered, periodic sequences; or (not shown) quantum algorithm kernels.~{\bf (d)} The results of running randomized mirror circuits of varied shapes on twelve quantum computers from IBM and Rigetti Computing (schematics show device layouts, dates when the experiments were performed). Each circuit's success probability $S$ is estimated from $\sim\!1000$ circuit repetitions and rescaled to $P= (S - 1/2^w)/(1-1/2^w)$ where $w$ is the circuit's width. The maximum, minimum and mean of $S$, over 40 circuits run at each width and depth, is shown for each device.  Green, black, and red lines (respectively) show the frontiers at which these statistics drop below $1/e \approx 0.37$. The maximum and minimum frontiers are calculated so that any discrepancy between them is statistically significant at $p=0.05$.}
\label{fig:1}
\end{figure*}

While several benchmarks have been proposed, few are efficiently verifiable. IBM's quantum volume benchmark \cite{cross2018validating}, like many application-derived benchmarks \cite{linke2017experimental, harrow2017quantum, wright2019benchmarking}, becomes infeasible to verify by classical simulation as the number of qubits grows. Google leveraged the extreme difficulty of verifying the results of their cross-entropy benchmarking circuits to demonstrate ``quantum supremacy'' \cite{arute2019quantum, neill2018blueprint, boixo2018characterizing}.  Other benchmarks present different problems.  For example, Clifford randomized benchmarking \cite{magesan2011scalable,emerson2007symmetrized,emerson2005scalable} uses a class of programs that, while efficiently verifiable, require so many gates when compiled on more than 3-5 qubits that they almost never run correctly on today's processors \cite{proctor2018direct}. Moreover, all of these benchmarks rely strictly on randomized, disordered programs. This limits their sensitivity to coherent noise \cite{kueng2015comparing}, and so they are unlikely to reflect the performance of structured programs that implement quantum algorithms.

We solve all of these problems by introducing a family of benchmarks that can probe the capability of any gate-model quantum processor --- including large ones capable of quantum advantage. To build these benchmarks, we begin with quantum circuits of varied sizes and structures that constitute challenging tasks for a processor.  Then we apply a procedure called “mirroring” \footnote{Our circuit mirroring technique is introduced in detail in Appendix~\ref{sec:mirroring}.} that transforms any circuit $C$ into a related suite $\{M_C\}$ of “mirror circuits” that are efficiently verifiable (see Fig.~\ref{fig:1}a).  Mirroring concatenates the original circuit $C$ with a quasi-inverse $\tilde{C}^{-1}$ that reverses $C$ up to a Pauli operation, and inserts special layers of operations before, after, and between $C$ and $\tilde{C}^{-1} $.  Quasi-inversion, inspired by the Loschmidt echo \cite{loschmidt1876uber} and early work on randomized benchmarking \cite{emerson2005scalable,emerson2007symmetrized}, ensures that each $M_C$ has a definite and easily verified target output, while the extra layers preserve the original circuit’s sensitivity to errors so that performance on $\{M_C\}$ faithfully represents performance on $C$. Unlike test circuits that yield high-entropy target distributions \cite{boixo2018characterizing,cross2018validating}, a mirror circuit’s performance is easily quantified by the probability $S$ of observing the ideal outcome. 

Mirror circuit benchmarks measure -- and inform prospective programmers about --- a processor’s capability to run specific programs (quantum circuits), rather than its ability to produce specific distributions \cite{arute2019quantum} or unitary transformations \cite{cross2018validating}.  The properties probed by such a benchmark are determined by the properties of the circuits in it. Mirror circuits can be efficiently constructed from circuits involving any number of qubits (circuit width, $w$) and logical cycles (circuit depth, $d$).  They can have any structure, enabling construction of benchmarks that serve as proxies for any quantum program. We built benchmarks from disordered (Fig.~\ref{fig:1}b) and highly structured (Fig.~\ref{fig:1}c) circuits, using gates that respect each processor’s connectivity, to probe different aspects of performance

We ran randomized mirror circuit benchmarks \footnote{A complete definition of randomized mirror circuits and the sampling distributions used in this experiment are given in Appendix~\ref{sec:rmcs}. Experimental details are given in Appendix~\ref{sec:b1}.} on twelve publicly accessible quantum computers from IBM \cite{ibmq2} and Rigetti Computing \cite{rigetti-qcs}. Their measured capabilities are displayed in Fig.~\ref{fig:1}d, using the framework of volumetric benchmarking \cite{blume2019volumetric}.  We probed each device at exponentially spaced ranges of circuit widths $w$ and benchmark depths $d$ \footnote{All circuits within a benchmark share a fixed $O(1)$ number of overhead layers; a mirror circuit’s benchmark depth counts only its non-overhead layers, and it equals twice the depth of the original circuit. This is explained further in the appendices.}, and for each width $w$ we tested several different embeddings of $w$ qubits into the available physical qubits. For each $d$, $w$, and embedding we ran 40 randomized mirror circuits.  For each shape $(w,d)$, Fig.~\ref{fig:1}d shows the best, worst, and average case polarization $P = (S -  \nicefrac{1}{2^w})/(1- \nicefrac{1}{2^w})$ for the best-performing $w$-qubit embedding. The polarization $P$ is a rescaling of the success probability $S$ that corrects for few-qubit effects. For example, $S =\nicefrac{1}{2}$ is reasonably good performance for a width-10 circuit ($P\simeq\nicefrac{1}{2}$) but represents total failure for a width-1 circuit ($P=0$). 

The volumetric benchmark plots \cite{blume2019volumetric} displayed in Fig.~\ref{fig:1}d provide an at-a-glance summary of these devices' capabilities to run random disordered circuits. They also encode considerable detail about the nature of the errors that limit capability. The mean polarization at each shape indicates the expected performance of a random circuit of that shape, and it is closely related to the fidelity of the logic gates (a standard measure of gate quality). The maximum and minimum polarization, $P_{\max}$ and $P_{\min}$, provide estimates of best- and worst-case capability, and their difference captures the variability --- how reliably do width and depth predict whether a random circuit will succeed? A large difference implies that whether a circuit can be successfully run on that processor depends not only on the circuit's shape, but also on its exact arrangement of gates, \ie, its structure. Our experiments reveal that certain processors' performance is strongly structure-dependent (\eg, Aspen 6) whereas other processors' performance is nearly structure-invariant (\eg, Vigo). This is highlighted by comparing the dotted lines in Fig.~\ref{fig:1}d that show the frontiers beyond which $P_{\mathrm{min}}$ (red) and $P_{\mathrm{max}}$ (green) fall below $1/e$ \footnote{The $\nicefrac{1}{e}$ threshold is arbitrary but convenient; under a simple, naive error model where each gate fails with probability $\epsilon \ll 1$, the frontier will include all circuits of size $\leq \nicefrac{1}{\epsilon}$. The maximum and minimum frontiers are calculated so that any discrepancy between them is statistically significant at $p=0.05$. The details of the statistical analysis are given in Appendix~\ref{sec:b1}.}. When a processor's performance is strongly structure-dependent, standard metrics derived from the average performance of random circuits \cite{boixo2018characterizing, cross2018validating, magesan2011scalable} will not reliably predict whether it can successfully run any particular randomly sampled circuit.

The success probability of a quantum circuit is dictated by a complex interplay between the structure present in that circuit and the structure of the errors. If errors are completely structureless (\ie, global depolarization), all circuits of a given shape will have the same success probability. But structureless errors are rare in quantum hardware. Error rates vary across qubits and noise is often correlated in time or space. Our results for randomized circuits provide a glimpse of this interplay. But random circuits are inefficient probes of structured errors \cite{kueng2015comparing}, because a typical randomized mirror circuit is almost completely disordered in space and time (Fig.~\ref{fig:1}b). To study the effects of structure, we can incorporate explicit long-range order, such as periodic arrangements of gates, into mirror circuits (Fig.~\ref{fig:1}c).  Periodic mirror circuits can be extremely sensitive to structured errors, supporting linear growth of coherent errors \cite{blume2016certifying} just as ordered lattice systems support ballistic transport of excitations~\cite{kohn1957quantum}. 

To investigate the interplay between circuit and error structures in real hardware, we benchmarked eight quantum processors using mirror circuits both with and without long-range order. We used periodic mirror circuits constructed by repeating a short unit cell of circuit layers (Fig.~\ref{fig:1}c) selected so that every circuit with $w >1$ had a two-qubit gate density of $\xi \approx \nicefrac{1}{8}$. Concurrently, we ran similar but randomized mirror circuits, sampled so that $\xi=\nicefrac{1}{8}$ in expectation. All circuits have $\xi \leq \nicefrac{1}{2}$, and deviations from $\xi = \nicefrac{1}{8}$ are small circuit-size effects. We sampled and ran 40 circuits of each type at a range of widths and depths, using the best qubits according to the manufacturer's published error rates \footnote{A complete definition of randomized mirror circuits, periodic mirror circuits, and the sampling distributions used in this experiment are given in Appendices~\ref{sec:rmcs} and~\ref{sec:pmcs}. Experimental details are given in Appendix~\ref{sec:b2}.}. Results for four representative devices \footnote{The results for all eight processors are included in Appendix~\ref{sec:b2}.} are summarized in Fig.~\ref{fig:2}.

\begin{figure}[t!]
\includegraphics[width=8.5cm]{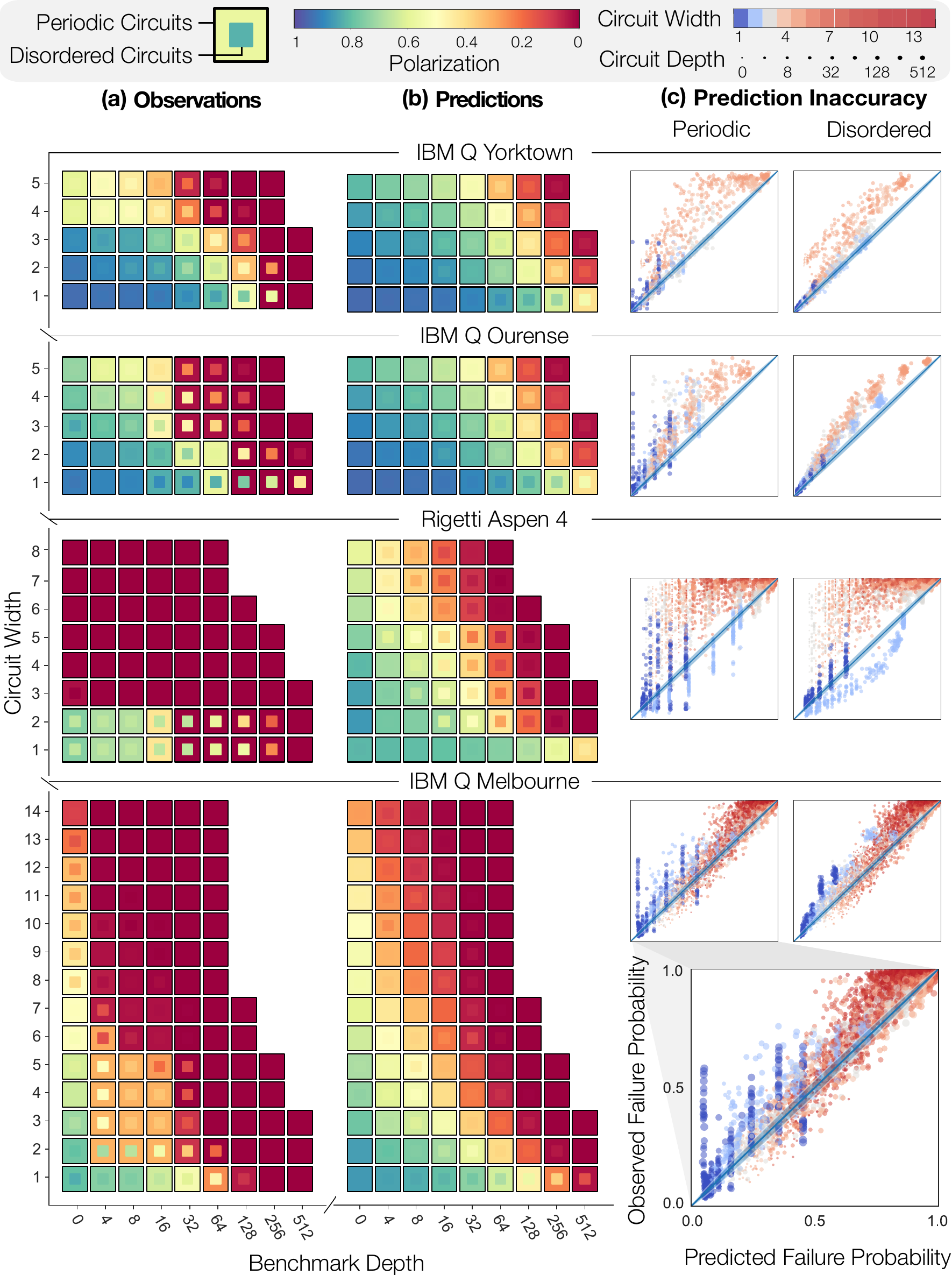}
\caption{{\bf Randomized benchmarks do not predict structured circuit performance.} Comparing the performance of quantum processors on circuits with and without long-range order.~{\bf (a)} The output polarization versus circuit width and depth for periodic (outer squares) and disordered (inner squares) circuits, minimized over all the test circuits that have that width or less and that depth or less.~{\bf (b)} Predictions from each device's error rates, accounting for the finite repeats of each circuit in the experiment  ($N=1024$) via a standard bootstrap.~{\bf (c)} The observed versus predicted failure rates for every circuit that was run. The blue diagonal bands are $2\sigma$ confidence regions: if the predictions were correct, $\sim$$95\%$ of the data would fall in them.} 
\label{fig:2}
\end{figure}

We found that the worst-case performance ($P_{\mathrm{min}}$) of periodic circuits was worse than that of disordered circuits for every processor, as shown in Fig.~\ref{fig:2}a. For some processors, the difference is dramatic --- \eg, Aspen 4 ran every width-1 disordered circuit up to depth $128$ successfully ($P \geq \nicefrac{1}{e}$), but failed on periodic circuits of depth 32. We conclude that testing a processor with disordered circuits cannot reliably predict whether that processor will be capable of running circuits with long-range order. Since circuits for quantum algorithms typically have long-range order, benchmarks like periodic mirror circuits are needed to predict the performance of algorithmic circuits \footnote{The structure in algorithm circuits can be reduced using a variety of randomization techniques \cite{wallman2015noise,campbell2017shorter}, but it is not possible to remove all structure in algorithm circuits.}.

We used our benchmarks to investigate one final question: can conventional error rates be used to predict a processor's capability? IBM and Rigetti publish an error rate for each logic operation (gates and readouts) in each device, updated every day after recalibration \cite{ibmq2,rigetti-qcs}. The presumption that these error rates can be used to accurately predict circuit success probabilities is the grounds for interpreting them as a measure of device quality. We recorded these error rates at the time of our experiments, and used them to predict the success probability for every circuit that we ran  \footnote{The method used to calculate these predictions is given in Appendix~\ref{sec:predictions}}. Fig.~\ref{fig:2}c shows a scatter plot comparing this prediction to experimental observations. In every case, the observed failure rates are dispersed widely around the prediction. This confirms the presence of unmodeled structure in the errors. The predictions are also biased towards over-optimism, suggesting the existence of significant error sources that are not captured by the error rates.  Comparing Figs.~\ref{fig:2}a and \ref{fig:2}b shows that, for every device, the observed worst-case performance is significantly worse than the performance predicted using published error rates. However, those error rates do not appear to be wrong --- they correctly predict the average performance of one- and two-qubit disordered circuits in most cases. Instead, we conclude that these discrepancies stem from unmodeled structure.  Structured errors affect structured and disordered circuits differently, and this cannot be captured by simple error rates.

The discrepancy between our observations and the predictions of the error rates reveals the types of structure present in the errors. All tested processors display performance that declines faster with circuit width than the error rates predict. This is a signature of crosstalk \cite{blume2019volumetric}. Similarly, the worst-case success rate of periodic circuits decays faster with depth than predicted, and than observed for disordered circuits. This is a signature of coherent errors \cite{blume2019volumetric,blume2016certifying}.  Mirror circuits with configurable structure are a simple tool for measuring the impact of these errors in large circuits like those needed for algorithms, so that they can be quantified and suppressed (\eg, with better calibrations) as necessary.

\begin{figure}[t!]
\includegraphics[width=8.5cm]{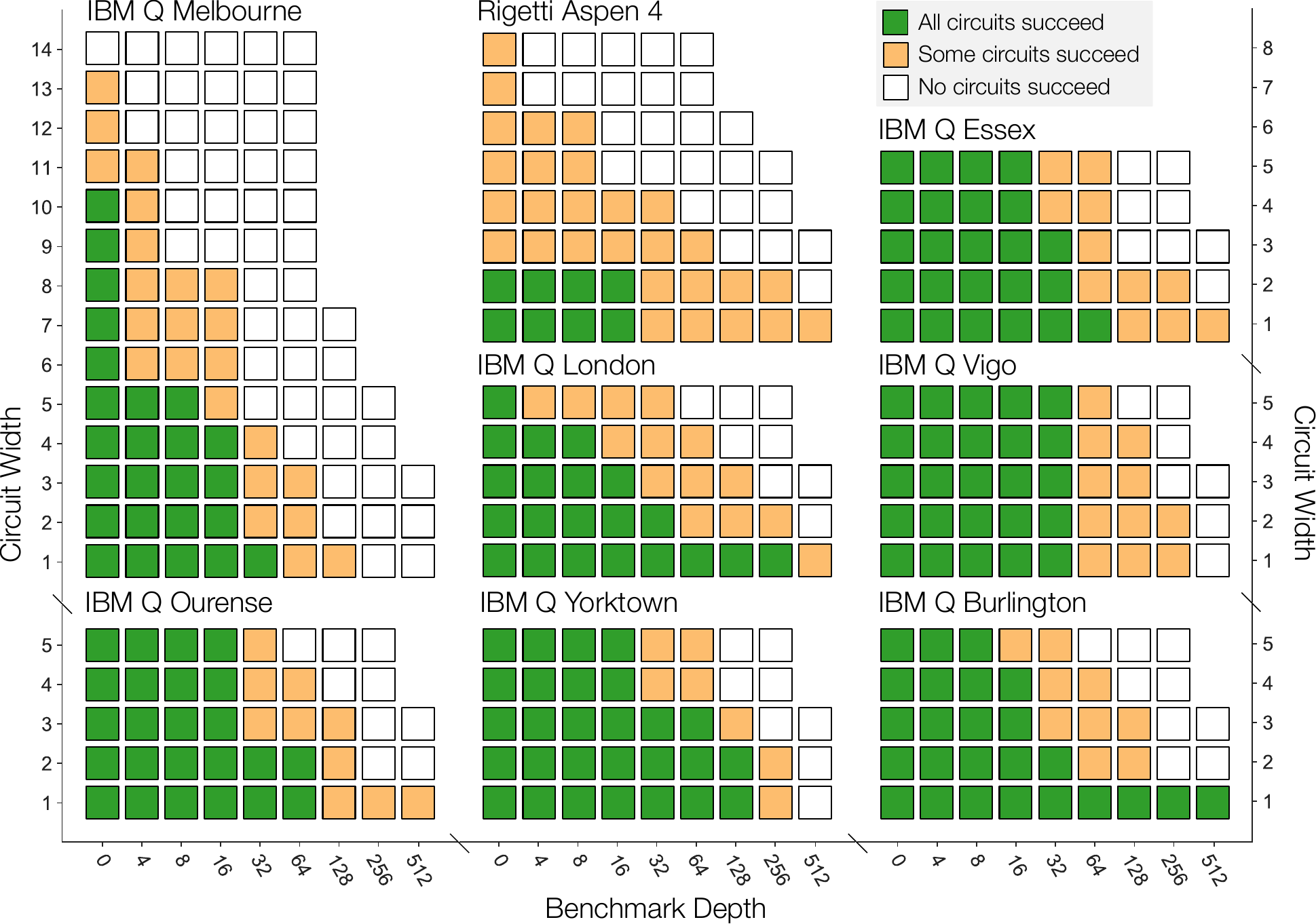}
\caption{{\bf Empirical capability regions.} The circuit shapes at which all (green), some (orange), and none (white) of the test circuits succeeded ($S \geq \nicefrac{1}{e}$). The test circuits have a two-qubit gate density of $\xi \approx \nicefrac{1}{8}$. If a target circuit with $\xi \approx \nicefrac{1}{8}$ lies in the green (white) region for a particular processor, then that processor will likely (not) execute the circuit successfully. Processor performance on circuits in the orange region is unpredictable.}
\label{fig:3}
\end{figure}

We have shown how to use mirror circuit benchmarks for detailed analysis of quantum processors' performance.  But our original goal was to capture performance in a simple and intuitive way.  So, in Fig.~\ref{fig:3}, we concisely summarize the performance of all eight devices tested with both kinds of mirror circuits, by dividing the circuit width $\times$ depth plane into ``success'', ``indeterminate'', and ``fail'' regions. They correspond to the circuit shapes at which (respectively) all, some, and none of the test circuits succeeded ($P \geq \nicefrac{1}{e}$). These empirical capability regions allow potential users to predict what circuits a processor is likely to be capable of running. A circuit whose shape falls into a processor's ``success'' or ``fail'' regions is likely to succeed or fail (respectively), because the test circuits probe both extremes of performance by including a variety of disordered and periodic circuits at each circuit shape.  Conversely, a processor's ability to successfully run a specific circuit whose shape falls within its ``indeterminate'' region depends unavoidably on that circuit's structure. Capability regions depend on two-qubit gate density ($\xi \approx \nicefrac{1}{8}$ in Fig.~\ref{fig:3}) and the threshold for success ($\nicefrac{1}{e}$ in Fig.~\ref{fig:3}), and can be easily adapted to particular applications by setting these parameters.

Quantum computational power is a double-edged sword.  The infeasibility of simulating quantum processors with 50+ qubits offers the possibility of computational speedups \cite{arute2019quantum, preskill2018quantum}, but simultaneously poses real problems for testing and assessing their capability.  As processors grow, users and computer engineers will need scalable, efficient and flexible benchmarks that can measure and communicate device capabilities.  Mirror circuit benchmarks demonstrate that this is possible, and highlight the scientific value of carefully designed benchmarks.

\bibliography{Bibliography}

*tjproct@sandia.gov

\begin{small}
\vspace{0.2cm}
\noindent \textbf{Acknowledgements:} This work was supported by the U.S. Department of Energy, Office of Science, Office of Advanced Scientific Computing Research through the Quantum Testbed Program and the Accelerated Research in Quantum Computing (ARQC) program, and the Laboratory Directed Research and Development program at Sandia National Laboratories. Sandia National Laboratories is a multi-program laboratory managed and operated by National Technology and Engineering Solutions of Sandia, LLC., a wholly owned subsidiary of Honeywell International, Inc., for the U.S. Department of Energy's National Nuclear Security Administration under contract DE-NA-0003525. All statements of fact, opinion or conclusions contained herein are those of the authors and should not be construed as representing the official views or policies of the U.S. Department of Energy, or the U.S. Government, or the views of IBM, or Rigetti Computing. We thank both the IBM Q and Rigetti Computing teams for extensive technical support, in particular Amy Brown, Jerry Chow, Jay Gambetta, Sebastian Hassinger, Ali Javadi, Francisco Jose Martin Fernandez, Peter Karalekas, Ryan Karle, Douglas McClure, David McKay, Paul Nation, Nicholas Ochem, Chris Osborn, Eric Peterson, Diego Moreda Rodriguez, Mark Skilbeck, Maddy Tod, and Chris Wood.
\end{small}

\section{Overview of the appendices}\label{sec:overview}
In the main text, we used \emph{mirror circuit benchmarks} to probe quantum computers' capabilities.  In these appendices:
 \begin{enumerate}
\item We explain why mirror circuits constitute a \emph{good} benchmark. 
\item We detail our experiments and data analysis.
\end{enumerate}
In this overview we explain what \emph{kind} of benchmark we seek to construct, we list desiderata for such a benchmark, and we provide a guide for the remainder of these appendices.

\subsection{The kind of benchmark we constructed}
Benchmarking a device means commanding it to perform a set of tasks, and measuring its performance on them.  The measured performance should be \emph{meaningful}. Prospective users should be able to extrapolate straightforwardly, from benchmark results, approximately how well the device would perform on \emph{their} use cases. But devices can be used in different ways, and for different tasks.  Distinct use cases require distinct benchmarks.  For example, a quantum computer can be commanded to (1) run a particular circuit, (2) apply a particular unitary, or (3) generate samples from a particular distribution. These task classes are categorically distinct, but each has real-world relevance.  Google, in their demonstration of quantum supremacy \cite{arute2019quantum}, benchmarked their Sycamore chip (against a supercomputer) by its performance at sampling a \emph{distribution}. IBM's quantum volume benchmark \cite{cross2018validating} challenges quantum processors to perform specific \emph{unitaries}, and cautions that it's cheating to sample from the resultant distribution without performing the specified unitary.  Randomized benchmarking \cite{knill2008randomized, magesan2011scalable, proctor2018direct} commands a processor to run specific \emph{circuits}, each one of which produces a trivial unitary and a trivial distribution. These illustrate three different ways that a quantum processor's task can be defined.

Here, we have adopted the third approach.  We benchmark processors by specifying concrete \emph{circuits}, not unitaries or distributions.  Therefore, these benchmarks measure a processor's ability to run circuits.  Their results should enable users to predict how well that processor will run \emph{other} circuits with similar properties.  Relative to the other paradigms mentioned above, this approach emphasizes the reliability of the processor's gates. Our paradigm isolates that aspect of performance from other properties, like qubit connectivity, gate set expressiveness, or the performance of a processor's classical compilation software.  Such benchmarks are and will be particularly useful to low-level quantum programmers who express their programs or algorithms as concrete circuits made of native gates, and then wish to predict how large a circuit can be run.  Benchmarks rooted in the other paradigms mentioned above are complementary, emphasizing other aspects of performance.  No single benchmark or paradigm is sufficient to capture all use cases.

\subsection{Desiderata for benchmark circuits}\label{sec:desiderata}
The specific benchmarks we use in the main text are particular cases produced by a general process. This process is designed to generate a set of circuits suitable for benchmarking from one or more exemplar circuits $C$ that represent a particular use case.  A good question to ask is ``If $C$ is a representative circuit, why not simply run $C$ itself as a benchmark?''  Doing this presents two problems.  

First, since the point of a benchmark is to measure performance, we \emph{must} be able to evaluate how well or accurately a given processor has run our benchmark circuits.  For many interesting and representative circuits, this is or will be impractical because good quantum algorithms can generate results that aren't classically simulable, and/or solve problems outside of NP (\ie, the result is not efficiently verifiable).  

Second, many circuits $C$ are intrinsically \emph{subroutines}, whose performance we wish to predict in contexts (\ie, within larger programs) that are \emph{a priori} unknown or only partially known.  A benchmark needs to run $C$ in context --- at a minimum, after state initialization and before measurement of all the qubits --- and a good benchmark must place it in \emph{representative} contexts, so that users can infer or predict how it is likely to perform in the specific context of \emph{their} use cases.  Even when $C$ is not a subroutine, but a full algorithm that defines its own context, the transformations required to make it easy-to-verify (solving the first problem above) can change that context, requiring additional work to ensure that $C$'s performance is probed in contexts that are representative of its original function.

To solve these problems, we need a process that transforms a user-specified circuit $C$ into a set of circuits or \emph{test suite} $\mathbb{S}(C)$, that can be run exhaustively or sampled from, and which satisfies the following key desiderata:
\begin{enumerate}
\item Even if $C$ is a subroutine that needs to be embedded into a larger circuit, every circuit in $\mathbb{S}(C)$ has a fully specified context including state initialization and measurement.
\item Each circuit in $\mathbb{S}(C)$ has a well-defined and easy to simulate \emph{target output}, which it would produce if implemented without errors, so that the performance of an imperfect implementation can be measured straightforwardly.
\item The success probabilities of the circuits in $\mathbb{S}(C)$ are \emph{representative} of how $C$ would perform in the context[s] where it might be used (which may be unknown).
\end{enumerate}

\subsection{Mirror circuit benchmarks}
We have developed a set of circuit transformations, collectively called \emph{mirroring}, that generate a set of benchmarking circuits from a user-specified circuit $C$, and that can be used to satisfy the above desiderata. These transformations generate \emph{mirror circuit benchmarks}. In our experiments we ran two particular types of mirror circuit benchmark: \emph{randomized mirror circuits} and \emph{periodic mirror circuits}. Appendices~\ref{sec:defs}-\ref{sec:pmcs} are dedicated to introducing these benchmarking methods:
\begin{itemize}
\item In Appendix~\ref{sec:defs} we introduce our notation and definitions, and review the background material required to present both our benchmarking methods and the theory supporting them.
\item In Appendix~\ref{sec:layer-set} we discuss the relative merits of defining benchmarking circuits over a standardized gate set versus over a gate set that is native to a particular processor, and we specify the approach that we take in our experiments.
\item In Appendix~\ref{sec:mirroring} we introducing our mirroring circuit transformations, and show how and why they satisfy the above desiderata.
\item In Appendix~\ref{sec:rmcs} we introduce randomized mirror circuits, and the particular types of randomized mirror circuits used in our experiments. 
\item In Appendix~\ref{sec:pmcs} we introduce periodic mirror circuits, and the particular type of periodic mirror circuits used in our experiments. 
\end{itemize}

Although Appendices~\ref{sec:defs}-\ref{sec:pmcs} discuss certain aspects of our experiments, they are primarily focused on describing our benchmarking methods in a general way that is applicable to almost any quantum computer. The final three Appendices focus on our particular experiments and the corresponding data analysis:
\begin{itemize}
\item In Appendix~\ref{sec:predictions} we explain how we used each processor's published error rates to predict the success probabilities of the mirror circuits that we ran.
\item In Appendix~\ref{sec:b1} we detail the randomized mirror circuit experiment, and the corresponding data analysis, that is summarized in Fig.~1d of the main text. We will refer to this as \emph{experiment \#1} throughout these appendices
\item In Appendix~\ref{sec:b2} we detail the randomized and periodic mirror circuits experiment, and the corresponding data analysis, that is summarized in Figs.~2 and 3 of the main text. We will refer to this as \emph{experiment \#2} throughout these appendices.
\end{itemize}
It is \emph{not} necessary to read these appendices in chronological order. Each appendix has been written to be as self contained as possible.

\section{Definitions}\label{sec:defs}
The purpose of this appendix is to define our notation and review the background material required throughout these appendices.

\subsection{Quantum circuits}
We use quantum circuits extensively in this paper, to define tasks and programs for quantum computers. Quantum circuits have been used so ubiquitously in the literature, for so many purposes, that it is difficult to define them in a simple yet universally valid way.  Broadly speaking, a quantum circuit \emph{describes} a (possibly complex) operation to be performed on a quantum computer, by specifying an arrangement of ``elementary'' operations (\eg, logic gates or subroutines) in sequence or in parallel, which if performed on the quantum computer will transform its state in a particular way. All the circuits that we consider in this paper can be represented, and implemented, as a series of \emph{layers}.

\subsubsection{Logic layers and unitaries}\label{sec:layers}
A $w$-qubit logic layer is an \emph{instruction} to apply physical operations that implement a particular unitary evolution on $w$ qubits. We denote the unitary corresponding to $L$ by $U(L) \in \text{SU}(2^w)$. Here $\text{SU}(2^w)$ denotes the $2^w$-dimensional special unitary group represented as matrices acting on the $2^w$-dimensional complex vector space $\mathcal{H}_{w}$ of pure $w$-qubit quantum states. It will also often be convenient to use the superoperator representation of a unitary, so we define $\mathcal{U}(L)$ to be the linear map
\begin{equation}
\mathcal{U}(L)[\rho] := U(L) \rho U(L)^{\dagger},
\end{equation}
where $\rho$ is a $w$-qubit density operator (a unit-trace positive semi-definite operator on $\mathcal{H}_{w}$), representing a general $w$-qubit quantum state.  We consider a logic layer $L$ to be entirely defined by the unitary $U(L)$, so --- by definition --- there is only one logic layer corresponding to each unitary. There will usually be many ways to implement a particular layer. Our methods are entirely agnostic as to how a layer is implemented, except that an attempt must be made to faithfully implement the unitary it defines. We use $L^{-1}$ to denote the logic layer satisfying 
\begin{equation}
U(L^{-1}) = U(L)^{-1}.
\end{equation}

There are two additional, special layers that can appear in our quantum circuits:  an \emph{initialization} or state preparation layer $I$ that initializes all qubits in the $\ket{0}$ state, and a \emph{readout} or measurement layer $R$ that reads out all qubits in the computational basis, producing a classical bit string and terminating the circuit.  Initialization can only appear as the first layer in a circuit, and readout can only appear as the last layer.  These layers are not unitary, and $U(\cdot)$ is not defined for them.

\subsubsection{Quantum circuits}\label{sec:circuits}
A quantum circuit $C$ over a $w$-qubit logic layer set $\mathbb{L}_w$ is a sequence of $d \geq 0$ logic layers that are all elements from $\mathbb{L}_w$. We will write this as
\begin{equation}
C = L_d L_{d-1} \cdots L_2 L_1,
\end{equation}
 where each $L_i \in \mathbb{L}_w$, and we use a convention where the circuit is read from right to left. The circuit $C$ is an instruction to applying its constituent logic layers, $L_1$, $L_2$, $\dots$, in sequence.  For the benchmarking purposes that we are concerned with in this paper, operations across multiple layers must \emph{not} be combined or compiled together by implementing a physical operation that enacts their composite unitary.  This notion of strict ``barriers'' between circuit layers is required in many benchmarking and characterization methods \cite{magesan2011scalable,magesan2012characterizing,blume2016certifying}, and we use it throughout this work. 
 
We consider two categories of quantum circuits, which have significantly different roles. \emph{Quantum input / quantum output} (QI/QO) circuits do \emph{not} use the initialization or readout layers.  \emph{Fixed input / classical output} (FI/CO) circuits begin with an initialization layer, and end with a readout layer.  There is a canonical mapping from QI/QO circuits to FI/CO circuits (by adding the initialization and readout layers) and back (by stripping them off). 

QI/QO circuits generally appear as subroutines.  A QI/QO circuit $C$ encodes a unitary map $U(C)$ on $w$ qubits given by
 \begin{equation}
U(C) =U(L_d) \cdots U(L_2) U(L_1).
 \end{equation}
FI/CO circuits represent complete, runnable quantum programs.  A FI/CO circuit $C$ encodes a probability distribution
 \begin{equation}
\mathsf{Pr}(x \mid C) = \left|\bra{x} U(L_d) \cdots U(L_2) U(L_1) \ket{0}^{\otimes w}\right|^2,
 \end{equation}
 over length-$w$ classical bit-strings, $x$.

\subsubsection{Circuit width, depth, size and shape}\label{sec:width-and-depth}
The circuit $C= L_dL_{d-1} \cdots L_2L_1$ defined over the $w$-qubit layer set $\mathbb{L}_w$ has
\begin{itemize}
\item a \emph{width} of $w$,
\item a \emph{depth} of $d$,
\item a \emph{size} of $w d$, and
\item a \emph{shape} of $(w,d)$.
\end{itemize}

The depth of a circuit is defined explicitly with respect to that circuit's specific layer set $\mathbb{L}_w$. Each specific quantum processor has a ``native'' layer set, generally corresponding to logic layers that can be implemented in a single unit of time.  For most processors, each native layer is some combination of one- and two-qubit gates in parallel. We do \emph{not} assume that every layer in the set $\mathbb{L}_w$ is native, nor that it can even be implemented with a short sequence of the native logic layers.  So implementing a circuit of depth $d$ could require many more than $d$ units of physical time.

Every (circuit paradigm) benchmark is defined by a set of circuits, and for every benchmark there is a set of ``overhead'' layers that are common to, and shared by, every circuit in the benchmark.  At a minimum, this overhead includes initialization and readout layers.  Therefore, in the context of a specific benchmark, we define three different depths for a circuit:
\begin{enumerate}
\item The \emph{full depth} $d_0$ of a circuit is the total number of layers, including initialization and readout, as defined above.
\item The \emph{benchmark depth} $d$ of a circuit is the total number of non-overhead layers, $d = d_0 - \mathrm{const}$, where the constant is the same for every circuit in a benchmark.
\item The \emph{physical depth} of a circuit is the total time taken to run a circuit assuming that every gate can be performed in single clock cycle (a single unit of time). Because this can depend strongly on hardware constraints, such as restrictions on parallelism, we do not use the physical depth in this work. 
\end{enumerate}
The circuit mirroring procedure that we discuss below typically adds overhead layers, and in our experiments there are five overhead layers (initialization, readout, and three extra logic layers).  In the main text, we report the benchmark depth defined by $d = d_0 - 5$. 

\subsubsection{Pauli layers}\label{sec:pls}
The $w$-qubit Pauli layers $\mathbb{P}_w$ are the $4^w$ logic layers that instruct the processor to implement $w$-fold tensor products of the four standard Pauli operators $I$, $X$, $Y$ and $Z$. For all $Q_1,Q_2 \in \mathbb{P}_w$,
\begin{equation}
\mathcal{U}(Q_2 Q_1) = \mathcal{U}(Q_3)
\end{equation}
for some $Q_3 \in \mathbb{P}_w$, \ie, $\mathcal{U}(\mathbb{P}_w)$ is a group, where $\mathcal{U}(\mathbb{L}) := \{\mathcal{U}(L)\}_{L \in \mathbb{L}}$ for any layer set $\mathbb{L}$. The Pauli operators induce bit flips and/or phase-flips on the qubits. So, if
\begin{equation}
\mathcal{U}(L_dL_{d-1} \cdots L_2L_1) = \mathcal{U}(Q)
\end{equation}
 for some Pauli layer $Q\in \mathbb{P}_w$, then the circuit $C = R L_dL_{d-1} \cdots L_2L_1 I$ will deterministically output a $w$-bit string that is specified by $Q$. This is a property that holds for all our benchmarking circuits. For any such circuit, its  \emph{target} bit string is the unique $w$-bit string that the circuit will output if it is implemented perfectly.

\subsubsection{Clifford layers and circuits}\label{sec:cls}
All the benchmarking circuits in our experiments contain only Clifford layers. A $w$-qubit logic layer $L$ is a Clifford layer if, for each $Q \in \mathbb{P}_w$,
\begin{equation}
\mathcal{U}\left(L Q L^{-1}\right)  = \mathcal{U}\left(Q'\right)
\end{equation}
for some Pauli layer $Q' \in \mathbb{P}_w$ \cite{aaronson2004improved}. Note that the Pauli layers are also Clifford layers, and $\mathcal{U}(\mathbb{C}_w)$ is a group where $\mathbb{C}_w$ denotes the set of all $w$-qubit Clifford layers. If a circuit contains only Clifford layers we refer to it as a Clifford circuit.

\subsection{Modeling quantum processors}\label{sec:markovian}
In these appendices we will show how a processor's performance on our mirror circuit benchmarks depends on the magnitude and type of the imperfections in that processor. Here we introduce our notation for modeling errors in quantum processors, and review the relevant standard definitions.

\subsubsection{The Markovian error model}
We will use $\Lambda(\cdot)$ to map from instructions --- layers or circuits --- to a mathematical object that models a processor's implementation of that instruction. In particular:
\begin{itemize}
\item  For a FI/CO circuit $C$, $\Lambda(C)$ is the distribution over $w$-bit strings that each run of $C$ on that processor is sampling from.
\item  For a QI/QO circuit $C$, $\Lambda(C)$ denotes a map from $w$-qubit quantum states to $w$-qubit quantum states. 
\end{itemize}
Our theory will use the \emph{Markovian error model} \cite{sarovar2019detecting} in which
\begin{itemize}
\item $\Lambda(I)$ is a fixed $w$-qubit density operator.
\item For any unitary logic layer $L$, $\Lambda(L)$ is a fixed completely positive and trace preserving (CPTP) linear map from $w$-qubit density operators to $w$-qubit density operators.
\item $\Lambda(R)$ is a positive-operator valued measure (POVM), \ie,
\begin{equation}
\Lambda(R)= \{\Lambda(R)_b\}_{b \in \mathbb{B}_w},
\end{equation}
 where $\mathbb{B}_w$ is the set of $w$-bit strings, the $\Lambda(R)_b$ are positive operators, and $\sum_b \Lambda(R)_b = \Id$.
\end{itemize}
The map implemented by a QI/QO circuit $C=L_d\cdots L_2L_1$ is then
\begin{equation}
\Lambda(C) = \Lambda(L_d) \cdots  \Lambda(L_2)\Lambda(L_1),
\end{equation}
where we have denoted composition of linear maps by multiplication (\ie, $\Lambda(L')\Lambda(L)$ represents the composition of the two linear maps). 

Similarly, for a FI/CO circuit $C=RL_d\cdots L_2L_1I$, $\Lambda(C)$ is a probability distribution over $w$-bit strings where the probability of the bit-string $b$ is
\begin{equation}
\Lambda(C)_{b} = \text{Tr}\left[ \Lambda(R)_b\Lambda(L_d) \cdots  \Lambda(L_1)[\Lambda(I)] \right],
\end{equation}
This error model can describe many common error modes in quantum processors --- including local coherent, stochastic and amplitude damping errors, as well as complex many-qubit errors like stochastic or coherent crosstalk \cite{gambetta2012characterization,sarovar2019detecting}. 

\subsubsection{Stochastic Pauli channels}\label{sec:pauli-channels}
Stochastic Pauli channels, and the special case of depolarizing channels, will have an important role in our theory of mirror circuit benchmarks.
A $w$-qubit stochastic Pauli channel is parameterized by a probability distribution over the $4^w$ Pauli operators: $\{\gamma_{Q}\}_{Q \in \mathbb{P}_w}$ with $\sum_{Q \in \mathbb{P}_w} \gamma_Q = 1$ and $\gamma_Q \geq 0$. The stochastic Pauli channel specified by $\{\gamma_{Q}\}$ has the action
\begin{equation}
\mathcal{E}_{\text{pauli},\{\gamma_{Q}\}}[\rho] := \sum_{Q \in \mathbb{P}_w} \gamma_Q U(Q) \rho U(Q)^{-1}. \label{eq:pauli-channel}
\end{equation}
A $w$-qubit depolarizing channel ($\mathcal{D}_{w,\epsilon}$) is a special case of a stochastic Pauli channel that is parameterized only by an error rate $\epsilon$:
\begin{equation}
\mathcal{D}_{w,\epsilon}[\rho] :=(1 - \epsilon) \rho + \frac{\epsilon}{4^w - 1} \sum_{Q \in \mathbb{P}_{w, \text{err.}}} U(Q) \rho U(Q)^{-1}, \label{eq:global-dep}
\end{equation}
where $\mathbb{P}_{w,\text{err.}}$ is the Pauli layers excluding the identity Pauli layer.

A $w$-qubit depolarizing channel is \emph{not} the $w$-fold tensor product of one-qubit depolarizing channels, that is, $\mathcal{D}_{w,\epsilon} \neq \mathcal{D}_{1,\epsilon'}^{\otimes w}$ for any $\epsilon'$ except for the special cases of the identity channel ($\epsilon=0$) and the maximally depolarizing channel ($\epsilon =(4^w-1)/4^w$). A $w$-qubit depolarizing channel induces highly correlated errors, whereas the $w$-fold tensor product of one-qubit depolarizing channels induces independent errors.

\subsubsection{Process fidelity}\label{sec:fidelity}
As we will show later, performance on our mirror circuit benchmarks have a relationship to the fidelity of the processor's implementation of the circuit[s] from which that benchmark was constructed, via ``mirroring.'' There are two commonly used definitions for the ``process fidelity'' --- the average fidelity and the entanglement fidelity. The average fidelity ($F_a$) of a $w$-qubit process $\mathcal{E}$ to the identity process is defined by \cite{nielsen2002simple}
\begin{equation}
F_{a}(\mathcal{E}) :=  \int d\psi \, \bra{\psi} \mathcal{E}[\ket{\psi}\bra{\psi}] \ket{\psi}, \label{eq:agf}
\end{equation}
where the integral is over the unique $\text{SU}(2^w)$-invariant measure on pure states. The entanglement fidelity ($F_{e}$) is defined by \cite{nielsen2002simple}
\begin{equation}
F_{e}(\mathcal{E}) := \bra{\psi_e} ( \mathcal{E} \otimes \mathcal{I})[\ket{\psi_e}\bra{\psi_e}] \ket{\psi_e}, \label{eq:agf}
\end{equation}
where $\mathcal{I}$ is the $w$-qubit identity superoperator (\ie, $\mathcal{I}[\rho]=\rho$), and $\ket{\psi_e}$ is any maximally entangled state in $\mathcal{H}_w\otimes \mathcal{H}_w$. The entanglement and average fidelity are related via \cite{nielsen2002simple}:
\begin{equation}
F_{e}(\mathcal{E}) = \left(1 + \nicefrac{1}{2^w}\right)F_a(\mathcal{E})  - \nicefrac{1}{2^w}. \label{eq:Fe-Fa}
\end{equation}
The average infidelity ($\epsilon_a$) and entanglement infidelity ($\epsilon_e$) are simply defined by
\begin{align}
\epsilon_{a}(\mathcal{E}) &:= 1 - F_{a}(\mathcal{E}),\label{eq:inf1}\\
\epsilon_{e}(\mathcal{E}) &:= 1 - F_{e}(\mathcal{E}).\label{eq:inf2}
\end{align}

Although the average fidelity is more widely used in the literature, for our purposes the entanglement fidelity is more relevant. This is because $F_e$ accounts for errors that are only apparent when a circuit is used as a subroutine inside a circuit on more qubits, whereas $F_a$ does not (note that $F_{e} < F_a$, unless $F_a=F_e=0$ or $1$). Therefore, this is the definition that we use for `the process [in]fidelity'. The entanglement infidelity of a stochastic Pauli channel has a simple and intuitive property: it is equal to the probability that the channel induces any Pauli error, \ie, 
\begin{equation}
\epsilon_{e}\left(\mathcal{E}_{\text{pauli},\{\gamma_{Q}\}}\right) = \sum_{Q \in \mathbb{P}_{w,\text{err.}}} \gamma_{Q}.
\end{equation}
In the special case of a depolarizing channel, $\epsilon_{e}(\mathcal{D}_{w, \epsilon}) = \epsilon$.

\section{Volumetric circuit benchmarks}\label{sec:layer-set}
The benchmarks constructed and deployed in this paper are examples of \emph{volumetric benchmarks} \cite{blume2019volumetric}. This means that each circuit in the benchmark has a well-defined width $w$ and depth $d$, that circuits with a range of $w$ and $d$ are selected, and that the data analysis sorts those circuits by $w$ and $d$.  Therefore, it is essential that the nature of these circuits \emph{and} the precise operational meaning of width and depth be stated clearly. A circuit's width is the number of qubits required to run it, and its depth is the number of layers that appear in it.  But both of these definitions are subject to non-obvious subtleties, especially depth. Depth is defined with respect to a particular set of logic operations (see Appendix~\ref{sec:width-and-depth}). Therefore, the benchmarking analysis depends critically on \emph{which} set of logic layers were used to define the benchmark circuits. The purpose of this appendix is to discuss several ways to choose layer sets, and then to describe the layer sets used in our experiments.

\subsection{Constructing layer sets from gate sets}
Layers are just instructions defining $w$-qubit unitary operations (see Appendix~\ref{sec:layers}). Many diverse layer sets could be defined for circuit benchmarks. For example, it is possible to define layers that perform very complicated unitaries that have to be compiled into complex circuits of one- and two-qubit gates. Conversely, it is possible to define layers that can be performed in a single clock cycle (on a specified processor).  The layer sets we use in this paper are composed of layers that are closer to the second example --- their ``physical depth'' (the number of clock cycles required for implementation) is relatively small.

In the layer sets used for our benchmarks, every allowed $w$-qubit layer is constructed by combining one- and two-qubit gates, chosen from a small gate set, in parallel.  Each of the $w$ qubits is acted on by at most one gate.  The gate set contains an \emph{idle} gate, and every qubit \emph{not} targeted by a nontrivial gate is said to be acted on by that idle gate.  By saying that individual gates are ``combined in parallel'', we are not saying that the processor has to implement them simultaneously.  Recall that a layer defines a unitary, not an implementation.  We are defining layers that \emph{could in principle} be implemented in parallel, within a single time step, by a processor that (1) can perform every gate in the gate set in a single time step, and (2) can perform them simultaneously.  But real processors are not required to do so --- the individual gates in a layer can be serialized and/or compiled into more elementary operations.

A precise description of how our layer sets are constructed from gate sets is as follows:
\begin{enumerate}
\item A $k$-qubit gate $G$ is an instruction to perform a specific unitary on $k$ qubits.  We only consider $k=1,2$.
\item A gate set $\mathbb{G} = \{G_1\ldots G_n\}$ is a list of 1- and 2-qubit gates.  Each gate could, in principle, be applied to any qubit (for 1-qubit gates) or any ordered pair of qubits (for 2-qubit gates).  However, connectivity constraints (see below) can be specified, and they restrict the qubits and/or ordered pairs of qubits on which a given gate can be applied.
\item We consider only gate sets that contain (a) exactly \emph{one} 2-qubit gate; (b) a 1-qubit ``idle gate''; and (c) any number of additional 1-qubit gates.
\item A $w$-qubit layer is constructed by assigning gates from $\mathbb{G} = \{G\}$ to specific qubits. In each $w$-qubit layer, each of the $w$ qubits is acted on by exactly one gate, which may be the idle gate.
\end{enumerate}

A $w$-qubit layer set $\mathbb{L}_w$ can be constructed, as above, by starting with a gate set and generating \emph{all} possible $w$-qubit layers of this form. We define smaller layer sets by allowing all and only those layers that respect:
\begin{enumerate}
\item \emph{A connectivity constraint} ($\Upsilon_{c}$) that specifies which qubits, or ordered pairs of qubits, each gate can be assigned to.  We call this a connectivity constraint because the most important type of assignment constraint is a limitation on which ordered pairs of qubits the two-qubit gate can be applied to. The connectivity constraint can be defined to respect a processor's directed connectivity graph, so that a two-qubit gate \emph{only} appears in layers if that processor can implement it natively. The (undirected) connectivity graphs for all twelve processors that we benchmarked are shown in Fig.~1d.
\item \emph{A parallelization constraint} ($\Upsilon_{p}$) that specifies which assigned gates are allowed to appear together in a layer. This can be used to respect a processor's limited ability to perform some gates in parallel, \eg, perhaps only a single two-qubit gate can be performed in a layer.
\end{enumerate}
Enforcing these constraints can reduce (or eliminate) the need for additional circuit compilation at run time. This can simplify further analyses of the benchmark results, such as estimation of per-gate error rates.

\subsection{Layer sets for benchmark circuits}
The procedure given above defines a canonical layer set for each ($\mathbb{G}$, $\Upsilon_{c}$, $\Upsilon_{p}$), which contains all the layers that can be built from $\mathbb{G}$ and are consistent with the constraints $\Upsilon_{c}$ and $\Upsilon_{p}$.  A benchmark's layer set determines two of its properties:
\begin{itemize}
\item The circuits that can be constructed and included in the benchmark.
\item How depth is defined and calculated for a given circuit.
\end{itemize}
The first property impacts what aspect of processor performance the benchmark measures, while the second impacts how that performance is quantified.  So the choice of layer set --- \ie, of $\mathbb{G}$, $\Upsilon_{c}$, and $\Upsilon_{p}$ --- is significant.

A standardized, architecture-blind layer set can be defined by making the $\Upsilon_{c}$ and $\Upsilon_{p}$ constraints trivial --- \ie, allowing \emph{all} gate assignments and placing no restrictions on parallelization --- and choosing a generic architecture-independent $\mathbb{G}$ such as \cnot plus all 24 single-qubit Clifford gates (or all single-qubit gates if non-Clifford gates are allowed.) At the other extreme, we can define an architecture-specific layer set by choosing $\mathbb{G}$, $\Upsilon_{c}$, and $\Upsilon_{p}$ to match the `native' layer set of a specific processor that is to be benchmarked. Both are viable, useful options. 

Benchmarking circuits defined over these two extreme layer sets, respectively, probe different properties of a processor.  Performance on benchmarks defined over native layer sets will correlate directly with the error rate of the native gates, and will not capture how ``useful'' those native gates are, or how much the processor is limited by connectivity or lack of parallelism. Conversely, benchmarks defined over a standardized layer set with no connectivity constraints will penalize processors with lower connectivity (relative to ``native layer'' benchmarks), because each two-qubit gate between qubits that are non-adjacent for a particular processor will have to be decomposed into a sequence of gates on adjacent qubits.  In principle, this can be a desirable property, because it is expected to capture performance on realistic algorithm circuits.  But it is also hard to calibrate.  Exactly \emph{how} a particular benchmark of this type penalizes lower connectivity will depend on the details of the benchmarking circuits.  Different algorithmic circuits are expected to incur different amounts of overhead (penalty) when embedded into a particular connectivity \cite{holmes2020impact,linke2017experimental}.  Capturing this behavior faithfully may require designing a different benchmark for each type of algorithm circuit.

\subsection{The layer sets for experiments \#1 and \#2}
In our experiments, we intentionally avoid the complexities of limited connectivity by using layer sets that respect a processor's connectivity graph (in contrast to, \eg, Refs.~\cite{linke2017experimental, cross2018validating}). In particular, we choose a layer set constructed from:
\begin{enumerate}
\item A gate set consisting of a processor's native two-qubit gate and a subset of the single-qubit Clifford group (see Appendice~\ref{sec:b1-gates} and~\ref{sec:b2-gates} for details). The native two-qubit gate for IBM Q processors is \cnot~\cite{ibmq2}, and for Rigetti processors it is \cphase~\cite{rigetti-qcs}. (Note that here ``native'' means the entangling gate exposed by the processor's interface, which may or may not correspond to the ``raw'' two-qubit gate implemented in hardware.)
\item A connectivity constraint corresponding to the processor's directed connectivity graph (see Fig.~1d for the undirected connectivity graphs for all twelve processors that we benchmarked.) Note that this means that specific width-$w$ benchmarking circuits cannot be constructed until we have chosen a subset of $w$ physical qubits on which to run them, because different subsets of qubits in a processor may have different connectivity graphs. We do not allow a disconnected subset of $w$ qubits to be chosen.
\item No parallelization constraint. A layer can contain active (\ie, non-idle) gates on all the qubits, regardless of whether the processor actually implements all those gates at the same time. This does \emph{not} mean that a processor necessarily actually runs the gates from a layer in parallel (for Rigetti's processors, it is our understanding that only one active gate is implemented at a time, so every layer is serialized \cite{rigetti-qcs}). It merely means that we define circuit depth with respect to a layer set with parallel gates. On a processor that serializes every $w$-qubit layer, the compiled circuit's \emph{physical} depth (number of clock cycles required) may be up to a factor of $w$ higher than the benchmark depth. (It may be less than $w$ because some gates could take zero time when serialized, \eg, an idle gate can be skipped.)
 \end{enumerate}
 
It could be argued that our choice for the layer set of each processor does not provide a ``fair'' comparison between the processors. For example, a processor will typically perform better on our benchmarks if the connections in the connectivity graph corresponding to the worst performing two-qubit gates are removed. This is a direct consequence of our decision to benchmark the full set of native operations of a processor. But no single choice of layer set can provide an uniquely ``fair'' comparison of two processors.  Processors are described by a complex set of performance characteristics, which can only be fully explored and compared by using multiple benchmarks.  Some should capture the limitations stemming from restricted connectivity, while others should not.  We anticipate that mirror circuit benchmarks \emph{will} be easily adapted to explore aspects of performance related to device connectivity (as other benchmarks already do \cite{linke2017experimental, cross2018validating}), but that is distinct and future work.

\subsection{Self-inverse layer sets}
A layer set $\mathbb{L}_w$ is \emph{self-inverse} if and only if $L^{-1} \in \mathbb{L}_w$ for all $L \in \mathbb{L}_w$.  All the benchmarks and layer sets that we construct and use in this paper are self-inverse (in particular, note that \cnot and \cphase are self-inverse gates). It is possible to construct layer sets $\mathbb{L}_w$ that (1) are \emph{not} self-inverse, and (2) include one or more layers whose inverse requires a very deep circuit (\ie, many layers in $\mathbb{L}_w$). But this has few or no practical consequences for applying the methods we present here --- in all the commmonly found native layer sets we are aware of, generating $U^{-1}$ requires approximately (and often exactly) the same circuit depth as generating $U$ for any unitary $U \in \text{SU}(2^w)$. Throughout the rest of these appendices we assume a self-inverse layer set without further comment.

\begin{figure}[t!]
\includegraphics[width=8.5cm]{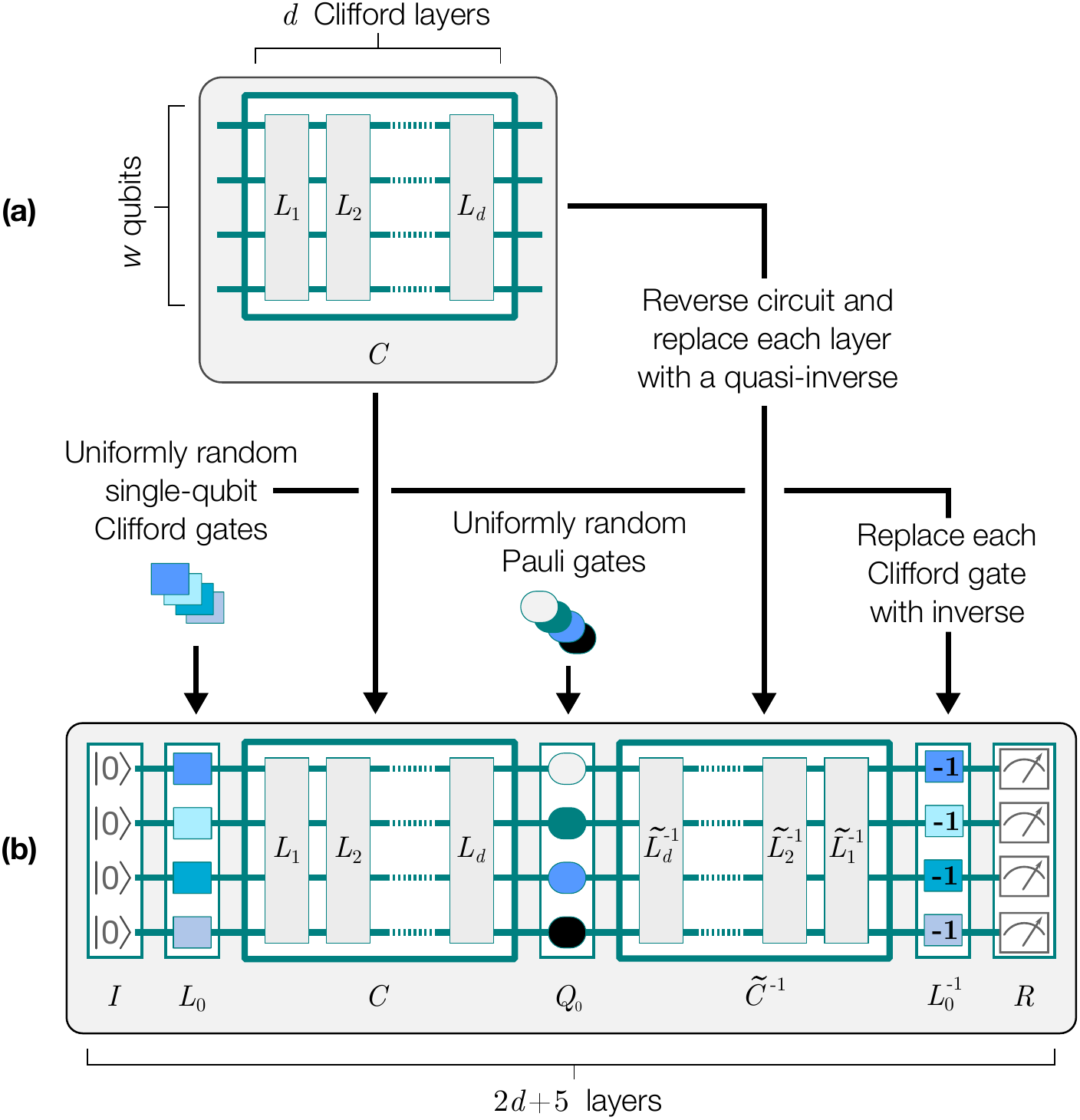}
\caption{\textbf{Transforming any Clifford circuit into mirror benchmarking circuits}. This figure illustrates our algorithm for transforming \textbf{(a)} any Clifford circuit $C$ into \textbf{(b)} a representative suite of benchmarking circuits $\mathbb{S}(C)$.  ``Representative'' means that a processor's average performance on circuits sampled at random from $\mathbb{S}(C)$ is representative of how well it could perform $C$, in a randomly chosen context (see text for details).  The \emph{quasi-inverse} layer $\tilde{L}_i^{-1}$ is the layer that inverts $L_i$ up to a particular Pauli operator $Q_i$ --- \ie, $\mathcal{U}(\tilde{L}_i^{-1} L_i) = \mathcal{U}(Q_i)$ --- where each $Q_i$ is drawn from a user-specified distribution (see text for details).  The circuits used in our experiments were generated via this algorithm, which is a specific case of the more general \emph{circuit mirroring} transformations that we discuss in Appendix~\ref{sec:mirroring}. These more general transformations can convert any circuit (not just Clifford circuits) into a benchmarking suite.}
\label{fig:clifford-mirroring}
\end{figure}

\section{Circuit mirroring}\label{sec:mirroring}
The mirror circuit benchmarks used in the main text were constructed using a set of circuit transformation procedures that we call, collectively, \emph{mirroring}. Mirroring transformations take arbitrary circuits, and create suites of benchmarking circuits that are closely related to the original circuit[s], but satisfy the benchmarking desiderata stated in Appendix~\ref{sec:desiderata} above. In this appendix, we introduce and motivate these transformations.  First, we summarize the \emph{specific} mirroring procedure used for the experiments we performed.  Then, we present each of the transformations that make up mirroring separately, because their utility extends beyond the specific procedure we used in this paper.

\subsection{Circuit mirroring as used in our experiments}
We refer to the specific circuit transformation used to generate the mirror circuit benchmarks used in our experiments as \emph{subroutine Clifford circuit} (SCC) mirroring.  SCC mirroring is illustrated in Fig.~\ref{fig:clifford-mirroring}.  A \emph{Clifford subroutine} is any  QI/QO circuit composed entirely of Clifford layers. SCC mirroring maps any Clifford subroutine to an ensemble $\mathbb{S}(C)$ of circuits that are suitable for benchmarking. SCC mirroring can be applied to FI/CO Clifford circuits, \ie, fully specified quantum programs composed of Clifford gates, by simply stripping away the program's initialization and readout layers. But, as we discuss below, SCC mirroring is designed to probe the performance of $C$ \emph{as a subroutine} --- \ie, with the expectation that it will not necessarily be applied to the $\ket{0}^{\otimes w}$ state, but to an arbitrary input state, generated in the context of a larger circuit that we do not know \emph{a priori}.  So SCC mirroring is not optimized to probe performance in the single FI/CO context, or any other specific context.

Given a Clifford QI/QO circuit $C = L_{d}L_{d-1} \cdots L_2 L_1$ defined over the layer set $\mathbb{L}_w$, the circuits in $\mathbb{S}(C)$ are defined as the following sequence of layers:
\begin{enumerate}
\item[(a)] The initialization layer $I$ that initializes all $w$ qubits to $\ket{0}$.
\item[(b)] A layer $L_{0}$ drawn from from $\mathbb{C}_{1}^w$ = \{all $24^w$ $w$-fold tensor products of the 24 single-qubit Clifford gates\}.
\item[(c)] The circuit  $C = L_d L_{d-1} \cdots L_2 L_1$.
\item[(d)] A layer $Q_0$ drawn from $\mathbb{P}_w$ = \{all $4^w$ $w$-qubit Pauli layers\}.
\item[(e)] The \emph{quasi-inversion} circuit
\begin{equation}
 \tilde{C}^{-1} = \tilde{L}_1^{-1}\tilde{L}_2^{-1} \cdots  \tilde{L}_{d-1}^{-1} \tilde{L}_{d}^{-1},
\end{equation}
where each \emph{quasi-inversion} layer $ \tilde{L}^{-1}_i$ is the unique circuit layer satisfying
\begin{equation}
\mathcal{U}(L_i \tilde{L}^{-1}_i) = \mathcal{U}(Q_i),
\end{equation}
 where $Q_i$ is a Pauli layer that is drawn from a user-specified distribution. (For example, this Pauli layer can be sampled from $\mathbb{P}_w$ uniformly and independently for each quasi-inversion layer. Alternatively, it can be set to the identity layer, so that each quasi-inversion layer is simply the inverse layer, \ie, $\tilde{L}^{-1} = L^{-1}$. We detail the choices made in our experiments later.)
\item[(f)] The layer $L_{0}^{-1} \in \mathbb{C}_{1}^w$ that inverts the Clifford layer $L_{0}$ performed in step (b).
\item[(g)] The readout layer $R$ that measures every qubit in the computational basis. 
\end{enumerate}
All circuits in $\mathbb{S}(C)$ therefore have the form:
\begin{equation}
    \mathbb{S}(C) = \left\{\,R\,L_0^{-1}\,\tilde{C}^{-1}\,Q_0\,C\,L_0\,I\, \right\}.
\end{equation}
The circuits in $\mathbb{S}(C)$ can be constructed by enumerating $L_{0}$ over the $24^w$ Clifford layers, $Q_0$ over the $4^w$ Pauli layers, and all other $Q_i$ according to the user-specified distribution.  More practically, they can be \emph{sampled} by drawing $L_{0}$ and $Q_0$ uniformly at random from those layer sets, and drawing each $Q_i$ from the given distribution.

Each circuit in $\mathbb{S}(C)$ is defined over the layer set $\mathbb{L}_w' = \mathbb{L}_w \cup \mathbb{P}_w \cup \mathbb{C}_{1}^w$ (where $\mathbb{A} \cup \mathbb{B}$ denotes the union of sets $\mathbb{A}$ and $\mathbb{B}$), and it has shape $(w, 2d+5)$. So if $C$'s original layer set $\mathbb{L}_w$ does not contain $\mathbb{P}_w$ and $\mathbb{C}_{1}^w$, then the circuits in $\mathbb{S}(C)$ are defined over a larger layer set than $C$.  This generally has no meaningful consequences, because those single-qubit Pauli and Clifford layers can almost always be implemented with shallow circuits over native layers, with relatively low error rates (at least compared with layers containing 2-qubit gates).  The generally negligible error rates of these ``extra'' layers motivate their exclusion from benchmark depth (see Appendix~\ref{sec:width-and-depth} above) --- we define the benchmarking depth in SCC mirroring as $d = d_0 - 5$.

SCC mirroring is motivated by the benchmarking desiderata that we presented in Appendix~\ref{sec:overview}.  So we will now demonstrate that it satisfies each of them.

\vspace{0.2cm} 
\textbf{The first requirement} is that each circuit in $\mathbb{S}(C)$ must have an entirely specified context --- \ie, it must be a complete, runnable quantum program.  SCC mirroring satisfies this by construction (because the $I$ and $R$ layers are explicitly included).

\vspace{0.2cm} 
\textbf{The second requirement} is that each circuit in $\mathbb{S}(C)$ must have a target output that is easy to compute on a conventional computer.  SCC mirroring also satisfies this requirement, although the explanation is a bit longer.  Any circuit $C_{\text{scc}} \in \mathbb{S}(C)$ has the form $  C_{\text{scc}} = R C_{\text{scc}}^{(0)} I$ where
\begin{align} 
C_{\text{scc}}^{(0)} = L_0^{-1} \tilde{L}_1^{-1} \tilde{L}_2^{-1} \cdots  \tilde{L}_{d}^{-1} Q_0 L_d \cdots L_2L_1 
\end{align}
is the central (QI/QO) part of the circuit, which we now show implements an easily computed Pauli operation. For any $w$-qubit Pauli layer $Q^{(1)} \in \mathbb{P}_w$ and any $L_i \in \mathbb{L}_w$,
\begin{align}
\mathcal{U}\left(\tilde{L}_{i}^{-1} Q^{(1)} L_i\right) &= \mathcal{U}\left(Q_i L_{i}^{-1} Q^{(1)} L_i\right), \\
    & = \mathcal{U}\left(Q_i Q^{(2)}\right),\\
    & = \mathcal{U}\left(Q^{(3)}\right),
\end{align}
for some $Q^{(2)},Q^{(3)} \in \mathbb{P}_w$. The second equality holds because the Pauli group is closed under conjugation by Clifford operations, and the last because the Pauli group is closed under multiplication. Therefore 
\begin{equation}
\mathcal{U}(C_{\text{scc}}^{(0)}) = \mathcal{U}(Q'),
\end{equation}
 for some Pauli layer $Q' \in \mathbb{P}_w$. This Pauli layer can be calculated efficiently in the circuit's size on a conventional computer using, \eg, the ``CHP'' code of Aaronson \cite{aaronson2004improved} (CHP can simulate large circuits over many thousands of qubits in less than a second on an ordinary laptop). Therefore, if performed without errors, each circuit in $\mathbb{S}(C)$ always produces a unique and deterministic bit string specified by that circuit's $Q'$. 

Since each circuit in $\mathbb{S}(C)$ has a unique target output, how well a given processor ran that circuit is easily quantified by its success probability ($S$). $S$ is just the probability of seeing the target bit string, and it can be estimated efficiently from data. In our data analysis we rescale $S$ to the \emph{polarization} $P = (S - \nicefrac{1}{2^w})/(1 - \nicefrac{1}{2^w})$, for the reasons discussed in the main text and in Appendix~\ref{sec:pol}.  But this is just a linear rescaling, which has no impact on the theory discussed here.  So in the rest of this appendix we will analyze $S$ instead of $P$.

\vspace{0.2cm} 
\textbf{The third requirement} (and the most subtle) is that the performance of the circuits in $\mathbb{S}(C)$ must be \emph{representative} of how $C$ would perform in the context[s] where it might be used.  This desideratum is what requires us to map $C$ to an \emph{ensemble} of circuits (rather than just a single circuit), and it therefore motivates each of the randomized elements in the procedure outlined above.  

To show that SCC mirroring satisfies the third desideratum, we represent a processor's imperfect implementation of the QI/QO circuit $C$ by a $w$-qubit superoperator $\Lambda(C)$ (see Appendix~\ref{sec:defs}). This superoperator can be written as
\begin{equation}
\Lambda(C)
= \mathcal{E}(C)\mathcal{U}(C),
\end{equation}
where $\mathcal{E}(C)$ is an error map.  If the processor can run $C$ perfectly, $\mathcal{E}(C)$ would be the identity superoperator $\mathcal{I}$.  As we will explain in the remainder of this appendix, SCC mirroring creates a test suite $\mathbb{S}(C)$ with the following properties:
\begin{enumerate}
\item For any error superoperator $\mathcal{E}(C) \neq \mathcal{I}$ there is a circuit in $\mathbb{S}(C)$ for which $S < 1$. That is, unless a processor can implement $C$ perfectly in all contexts, there is at least one circuit in $\mathbb{S}(C)$ that will bear witness to the error.
\item The \emph{expected} value of $S$ for a circuit sampled from $\mathbb{S}(C)$ is closely related to the process fidelity of $\mathcal{E}(C)$. Therefore, the expected value of $S$ is approximately probing the performance of a processor on $C$ in a uniformly random context. We make this statement more precise later in this appendix.
\end{enumerate}

These two properties are a well-motivated sense in which a processor's performance on a set of benchmarking circuits derived from $C$ can be representative of the processor's performance on $C$. But it is not the only well-motivated interpretation of ``representative performance''.  SCC mirroring creates a benchmark whose \emph{average} performance is closely related to the \emph{average} fidelity with which the processor implements $C$.  A benchmark that captured the processor's \emph{worst-case} performance on $C$ --- \ie, the maximum probability, over all possible contexts, of getting the wrong output from running $C$ in that context --- would arguably be even more desirable.  But no benchmark can extract this information efficiently in $w$, because there are $e^{O(w)}$ possible contexts (\eg, input states).  Capturing worst-case performance, without additional prior information, requires exhaustively exploring all of those contexts, which is infeasible.  So the notion of ``representative performance'' achieved by SCC mirroring is not unique, but it is both natural and achievable.

The remainder of this appendix presents the collection of circuit transformations that, together, constitute \emph{mirroring}.  Combined in a specific way, they generate the SCC mirroring procedure explained above. Since all of the experiments we report in the main text use SCC mirroring exclusively, our primary aim is to prove that SCC mirroring satisfies the two properties stated above. But the mirroring transformations listed here are more powerful.  They can also be used to generate (1) benchmarking circuits with different properties, and (2) benchmarks from non-Clifford circuits. So a secondary aim of this appendix is to explain the transformations independently, and illustrate this extensibility.

\subsection{Transformation 1: simple circuit mirroring}\label{sec:simple-mirroring}
Many classical programs have a unique ``right'' answer, which makes it easy to detect (and benchmark) errors in classical computers.  But interesting quantum circuits don't generally produce definite outcomes (\ie, a unique bit string) even when run without errors.  Instead, the post-measurement outcome of generic quantum programs is a high-entropy \emph{distribution} over bit strings, and it can be extremely costly to verify that this distribution matches the target, \ie, that the \emph{right} distribution is being produced. So to enable benchmarks derived from generic circuits, the first thing we need is a way of transforming interesting quantum circuits so that they \emph{do} produce definite outcomes.  The rather obvious solution is time reversal, and we call the particular transformation that we use \emph{simple circuit mirroring}. This transformation turns any circuit into a definite-outcome circuit, satisfying our second requirement, at the cost of creating some new problems that we will address later.

Simple circuit mirroring is essentially a type of Loschmidt echo \cite{loschmidt1876uber}. It maps \emph{any} shape $(w,d)$ QI/QO circuit $C = L_d \cdots L_2L_1$ over some self-inverse layer set $\mathbb{L}_w$ into a single shape $(w,2d+2)$ FI/CO circuit $M(C)$ over $\mathbb{L}_w$,
\begin{equation}
    M(C) = R\,C^{-1}\,C\,I,
\end{equation}
consisting of:
\begin{enumerate}[label=(\roman*)]
\item The initialization layer $I$ that initializes all $w$ qubits to $\ket{0}$.
\item The circuit $C= L_dL_{d-1} \cdots L_2L_1$.
\item The \emph{inversion circuit}
\begin{equation}
 C^{-1} = L_1^{-1} L_2^{-1} \cdots L_{d-1}^{-1}L_{d}^{-1},
\end{equation}
consisting of the layers of $C$ in the reverse order and with each layer $L$ replaced with its inverse $L^{-1}$.
\item The readout layer $R$ that measures every qubit in the computational basis. 
\end{enumerate}

The inversion circuit $C^{-1}$ implements the inverse unitary to $C$, \ie,
\begin{equation}
U(C)U(C^{-1}) = \Id.
\end{equation}
Therefore, for any circuit $C$, if $M(C)$ is performed without error, it will deterministically return the all-zeros bit string. Simple circuit mirroring achieves the first two of our three desiderata for a circuit transformation (see above) for generating a benchmarking suite $\mathbb{S}(C)$ from a circuit $C$: the single-element set $\mathbb{S}_{sm}(C) = \{M(C)\}$ generated by simple circuit mirroring contains a single circuit with an entirely specified context (it is a complete program) and an efficiently simulable target output (it is the all-zeros bit string). 

Simple circuit mirroring is a good starting point for satisfying the third desiderata, but, unaltered, it does not meet it. The circuit suite $\mathbb{S}_{sm}(C) = \{M(C)\}$ generated by simple circuit mirroring is not representative of $C$ in \emph{any} meaningful sense. (Unless strong assumptions are made about the types of errors that a processor is subject to, $M(C)$ is only representative of $C$ in the trivial sense that a processor's performance on $M(C)$ is representative of its performance on $C$ in the context of inserting $C$ into that simple circuit mirror circuit.) The limitations of simple circuit mirroring all stem from the fact that it involves running $C$ in a single context. Three specific effects that limit the usefulness of $\mathbb{S}_{sm}(C)$ are:

\begin{enumerate}
\item \emph{Systematic error cancellation}. In simple circuit mirroring, the circuit $C$ is always followed by the circuit $C^{-1}$. This means that it is possible for systematic (coherent) errors in the implementation of $C$ to exactly cancel with systematic errors in the implementation of $C^{-1}$. For example, if $\Lambda(C) = \mathcal{V}$ and  $\Lambda(C^{-1}) = \mathcal{V}^{-1}$ for some unitary superoperator $\mathcal{V}$ then $S=1$, up to contributions from errors in qubit initialization and readout. This does \emph{not} require that $\Lambda(C)$ is even close to the target evolution $\mathcal{U}(C)$. This is a well-known effect with the Loschmidt echo, which tests whether an evolution can be reversed, not whether a desired evolution can be implemented accurately.
\item \emph{A single input state}. The state input into $C$ is always $\ket{0}^w$, so simple circuit mirroring is insensitive to any errors that do not impact $\ket{0}^w$.
\item \emph{A single measurement basis}. The measurement is always in the computational basis, so simple circuit mirroring is insensitive to any errors that, once commuted through the circuit, manifest as errors that have no observable impact after projection onto $\bra{0}^w$ (such as dephasing or coherent $\hat{z}$-axis errors).
\end{enumerate} 
 The three additional circuit transformations tools that we introduce below can be used to place $C$ in a wider range of contexts. They start from $\mathbb{S}_{sm}(C)$ and map it to an altered and (typically) enlarged benchmarking suite. It is convenient to think of these three tools as a set of three configurable circuit transformation that are applied in order.

\subsection{Transformation 2: inserting a central subroutine} \label{sec:central-subroutine}
The first weakness of simple circuit mirroring, highlighted above, is that it can hide errors in $C$, because errors in $C^{-1}$ might systematically cancel out errors in $C$.  To solve this problem, we introduce another transformation called \emph{central subroutine insertion}, which we apply to the test suite $\mathbb{S}_{sm}(C) = \{M(C)\}$ obtained from simple circuit mirroring.  It constitutes inserting each of a set $\mathbb{A}$ of subroutines --- \ie, QI/QO circuits --- between $C$ and $C^{-1}$. This transformation acts on $\mathbb{S}_{sm}(C)$ as:
\begin{equation}
 \{ M(C) = R C^{-1} C I \} \to \{ M_A(C) = R C^{-1} A C  I \}_{A \in \mathbb{A}}.
\end{equation}
Central subroutine insertion generates a larger circuit suite,
\begin{equation}
\mathbb{S}_{\mathbb{A}}(C) = \{M_A(C) \}_{A \in \mathbb{A}},
\end{equation}
that can be run exhaustively or sampled from.
The point of the central subroutine is to prevent systematic errors in $C$ and $C^{-1}$ from canceling each other.  It only works if $\mathbb{A}$ is chosen carefully, to satisfy three competing criteria:
\begin{enumerate}
\item The subroutines in $\mathbb{A}$ should be sufficiently diverse that no possible error mode on $C^{-1}$ can systematically cancel out errors on $C$ in \emph{every} circuit in $\mathbb{S}_{\mathbb{A}}(C)$. As an obvious example, an $\mathbb{A}$ containing only the trivial, depth-0 circuit would not be sufficiently diverse.
\item Each circuit in $\mathbb{S}_{\mathbb{A}}(C)$, when run without error, should output a single, efficiently calculable bit string.  In some scenarios, achieving this requirement will require an additional transformation, as explained in Transformation 3 below.
\item The subroutines in $\mathbb{A}$ should be implementable with shallow circuits, so that running the circuits in $\mathbb{S}_{\mathbb{A}}(C)$ is not much harder than running simple mirror circuits.
\end{enumerate}

To make the ``sufficiently diverse'' condition above precise, we consider the linear map $\mathscr{L}_{C, \Lambda}$ on $w$-qubit superoperators (a so-called \emph{super-duper-operator}~\cite{crooks2008quantum}) defined by:
\begin{equation}
\mathscr{L}_{C,\Lambda}(\mathcal{S}) = \Lambda(C^{-1}) \mathcal{S} \Lambda(C).
\end{equation}
This map is parameterized by (1) a circuit $C$, and (2) a processor's $\Lambda(\cdot)$ map. We say that a processor implements a circuit $C$ perfectly if and only if $\Lambda(C) = \mathcal{U}(C)$.  Therefore, a processor perfectly implements both $C$ and $C^{-1}$ if and only if, for \emph{every} superoperator $\mathcal{S}$,
\begin{equation}
\mathscr{L}_{C,\Lambda}(\mathcal{S}) = \mathscr{L}_{C,\mathcal{U}}(\mathcal{S}).\label{eq:Ll=Lu}
\end{equation}
Simple circuit mirroring cannot tell us whether this is the case.  It only tells us about $\mathscr{L}_{C,\Lambda}(\mathcal{I})$, where $\mathcal{I}$ is the identity superoperator, because the processor's implementation of the QI/QO component of the simple mirror circuit $M(C)$ is
\begin{equation}
 \Lambda(C^{-1}C) = \mathscr{L}_{C,\Lambda}(\mathcal{I}).
\end{equation}
So simple circuit mirroring cannot be sensitive to all possible errors in $\Lambda(C)$ and $\Lambda(C^{-1})$, because Eq.~\eqref{eq:Ll=Lu} might hold for $\mathcal{S}=\mathcal{I}$, but not for all $\mathcal{S}$.

We can use $\mathscr{L}_{C,\Lambda}$ to more precisely state the first of our criteria for $\mathbb{A}$, introduced above. For any $\Lambda(C)$ and $\Lambda(C^{-1})$ superoperators for which $\Lambda(C) \neq \mathcal{U}(C)$ and/or $\Lambda(C^{-1}) \neq \mathcal{U}(C^{-1})$, there must exist an $A\in \mathbb{A}$ such that 
\begin{equation}
\mathscr{L}_{C,\Lambda}(\mathcal{U}(A)) \neq \mathscr{L}_{C,\mathcal{U}}(\mathcal{U}(A)).
\end{equation}
Without assumptions about the constituent superoperators, this holds if and only if $\mathcal{U}(\mathbb{A})=\{\mathcal{U}(A)\}_{A\in\mathbb{A}}$ spans the vector space of $w$-qubit superoperators. (Because $\Lambda(C)$ and $\Lambda(C^{-1})$ must be completely positive and trace preserving maps there are interesting edge cases where we can learn everything about $\Lambda(C)$ and $\Lambda(C^{-1})$ with a smaller set $\mathbb{A}$.) Ideally, $\mathcal{U}(\mathbb{A})$ should span that space \emph{uniformly} (as does, \eg, an orthonormal basis) to maximize sensitivity to all possible errors.  However, constructing a set of circuits that span the superoperator space requires nontrivial circuits.  So in SCC mirroring, we settle for a slight weaker (but much simpler) construction that detects \emph{almost} all errors.  

We choose an $\mathbb{A}$ containing all the $w$-qubit Pauli layers $\mathbb{P}_w$ (see Fig.~\ref{fig:clifford-mirroring}).  When $C$ is a Clifford circuit as in the main text (we address non-Clifford circuits briefly in the next subsection of this appendix), the Pauli layers are a particularly powerful choice for the following reasons.
\begin{enumerate}
\item \emph{Sensitivity to all small errors}. The Pauli group $\mathcal{U}(\mathbb{P}_w)$ does \emph{not} span superoperator space (there are only $4^w$ elements in $\mathcal{U}(\mathbb{P}_w)$, but superoperator space has dimension $4^w \times 4^w = 16^w$), but it has a property that is almost as good in this context. If $\mathscr{L}_{C,\Lambda} (Q) = \mathscr{L}_{C,\mathcal{U}}(Q)$ for all $Q\in \mathbb{P}_w$ then this implies that $\Lambda(C) =\mathcal{U}(QC)$ and $\Lambda(C^{-1}) =\mathcal{U}(C^{-1}Q)$ for some Pauli layer $Q$ that is the same in both equations, \ie, the correct unitaries are implemented up to multiplication by some Pauli operator. So the only errors that go undetected by $\mathbb{A}=\mathbb{P}_w$ are large, discrete, and unlikely except in an adversarial context.
\item \emph{Faithfulness in infidelity}. More than merely making all small errors \emph{detectable}, the Pauli group construction ensures that their average impact on the benchmark circuits faithfully reflects their impact in $C$ and $C^{-1}$.  If we average uniformly over $\mathbb{A}$, the effect of inserting a random Pauli layer between $C$ and $C^{-1}$ is to perform a Pauli twirl on the error maps for $C$ and $C^{-1}$, reducing them to stochastic Pauli channels \cite{knill2005quantum, wallman2015noise,ware2018experimental}.  For small errors, this ensures that the fidelity of the full benchmark circuit is very close to the product of the fidelities of $C$ and $C^{-1}$.
\item \emph{Efficiently calculable target outputs}. For any Pauli layer $Q$, $ \mathcal{U}(C^{-1} Q C) = \mathcal{U}(Q'),$
for some Pauli layer $Q'$. So $M_Q(C)$ will always output a single, efficiently calculable bit string determined by $Q'$,  if implemented perfectly.
\item  \emph{Unbiased target outputs}. Uniform sampling from $\mathbb{S}_{\mathbb{P}_w}(C)$ ensures that the target bit string is uniformly random, so biased readout errors cannot artificially boost or suppress the success probabilities $S$ of the circuits in $\mathbb{S}_{\mathbb{P}_w}(C)$ (again, on average).
\item \emph{Low-depth circuits.} Any Pauli layer can be implemented with a low-depth circuit over the native layer-set of a typical processor.
\end{enumerate}
   
 In the remainder of this appendix we will consider only the case of $\mathbb{A}=\mathbb{P}_w$.

\subsection{Transformation 3: replacing the inversion circuit with a suite of quasi-inversion circuits}
The reason that errors can systematically cancel in simple circuit mirroring is that $C$ is always followed by the same circuit, $C^{-1}$. Inserting a central subroutine prevents this error cancellation, but we can also reduce the correlation between layers in a mirror circuit by replacing $C^{-1}$ with a \emph{quasi-inversion} circuit $\tilde{C}^{-1}$. For a Clifford circuit $C$, this transformation maps each $M_Q(C)$ circuit to a set of circuits where the inverse circuit $C^{-1}$ has been replaced by each of a set of \emph{quasi-inversion} subroutines $\mathbb{Q}$. It is the map:
\begin{equation}
M_Q(C) = RC^{-1}QCI \to  \{M_{Q,\tilde{C}}(C) = R \tilde{C}^{-1} Q C I\}_{\tilde{C}^{-1} \in \mathbb{Q}},
 \end{equation}
 where $\mathbb{Q}(C)$ consists of all circuits of the form
 \begin{equation}
\tilde{C}^{-1} =\tilde{L}^{-1}_1 \tilde{L}_2^{-1} \cdots \tilde{L}_{d-1}^{-1} \tilde{L}_{d}^{-1}.
\end{equation}
Here each $\tilde{L}_{i}^{-1}$ runs over some set of $L$-dependent layers $\mathbb{Q}_1(L)$ that all implement unitaries that are equivalent to $\mathcal{U}(L^{-1})$ up to multiplication by a Pauli operator.

Different choices for $\mathbb{Q}_1$ result in different transformations. In our benchmarking experiments we use two transformations: the trivial transformation given by $\mathbb{Q}_1(L) = \{L^{-1}\}$, and the transformation in which $\mathbb{Q}_1(L)$ consists of all $4^w$ layers $\tilde{L}^{-1}$ that satisfy $\mathcal{U}(\tilde{L}^{-1}) = \mathcal{U}(\tilde{Q}'L^{-1})$ for some $\tilde{Q}' \in \mathbb{P}_w$ (which is similar to Pauli frame randomization \cite{knill2005quantum, wallman2015noise, ware2018experimental}). 

A similar transformation can be used to create mirror circuits with a central Pauli subroutine from non-Clifford circuits: in that case we choose the quasi-inverse layers as a function of the central Pauli layer $Q$, which allows us to construct quasi-inverse circuits for which the entire circuit implements a Pauli operator. As we do not use non-Clifford circuits in our experiments, we leave further details of this technique to future work.

\subsection{Transformation 4: inserting preparation and measurement subroutines}
The last of our circuit transformations is intended to address the last two limitations of simple circuit mirroring listed at the end of Appendix~\ref{sec:simple-mirroring}: that only the $\ket{0}^w$ state is input into $C$, and that readout is always in the computational basis. These limitations mean that any errors in the implementation of $C$ that do not affect the $\ket{0}^w$ state do not contribute to the failure rate of a simple mirror circuit, nor to the failure rates of the circuits in the expanded suites obtained from Transformations 2 and 3. If the circuit $C$ will only ever be applied to $\ket{0}^w$, then this is not a flaw as it represents the desired context. But to capture any other contexts, we need to implement additional input states and measurement bases so that performance on the benchmarking suite is representative of ability to perform $C$ in generic contexts. We do this by inserting ``fiducial'' \cite{blume2016certifying} subroutines just after initialization and before readout, respectively.

This procedure is parameterized by a set of QI/QO circuits $\mathbb{F}$ and it maps each circuit $M_{Q,\tilde{C}}(C) = R \tilde{C}^{-1} Q C I$ to a test suite
\begin{equation}
M_{Q,\tilde{C}}(C) \to  \{M_{Q,\tilde{C},F}(C) = R F^{-1} \tilde{C}^{-1} Q C F I\}_{F\in \mathbb{F}}.
 \end{equation}
Together, the four transformations generate the circuit suite
\begin{equation}
\mathbb{S}_{\mathbb{P}_w, \mathbb{Q}, \mathbb{F}}(C) = \{M_{Q,\tilde{C},F}(C) \}_{Q \in \mathbb{P}_w, \tilde{C}^{-1} \in \mathbb{Q}(C), F \in \mathbb{F}},
\end{equation}
which can be run exhaustively or sampled from. We need to choose $\mathbb{F}$ to satisfy the four competing criteria:
\begin{enumerate}
\item Each circuit in $\mathbb{S}_{\mathbb{P}_w, \mathbb{Q}, \mathbb{F}}(C)$ should still have an efficiently calculable target bit-string.
\item The fiducial subroutines should be implementable with shallow circuits over a typical processor's native gate, so that errors in these subroutines do not dominate the failure rate of the circuits in $\mathbb{S}_{\mathbb{P}_w,\mathbb{Q}, \mathbb{F}}(C)$, except perhaps for very shallow $C$.
\item The fiducials should be sufficiently diverse that they reveal all errors that are visible in any of the contexts in which $C$ will be used. In the case of a subroutine $C$ that is to be used in an \emph{a priori} entirely unknown context, this means that if $\Lambda( \tilde{C}^{-1} Q C) \neq \mathcal{U}(\tilde{C}^{-1} Q C)$ then there should be at least one $F \in \mathbb{F}$ for which $S < 1$ for the corresponding circuit. 
\item (Stretch goal) The fiducials should generate circuits that are \emph{uniformly sensitive} to all possible errors in $\Lambda( \tilde{C}^{-1} Q C)$, ensuring that the average performance over randomly sampled fiducials is closely related to the process fidelity of  $\Lambda( \tilde{C}^{-1} Q C)$.
\end{enumerate}

The first criterion is achieved by setting $\mathbb{F}$ to any subset of the $w$-qubit Clifford layers $\mathbb{C}_w$. The third criterion is satisfied if and only if $\mathbb{F}$ is informationally complete (\ie, it's element are sufficient for process tomography).  The fourth criterion is achieved by a set $\mathbb{F}$ that generates a 2-design, such as the $w$-qubit stabilizer states, which is achieved by the full $w$-qubit Clifford layer set $\mathbb{C}_w$ \cite{gross2007evenly, dankert2009exact}.  But the elements of $\mathbb{C}_w$ cannot be implemented with $O(1)$ depth circuits, so the full $w$-qubit Clifford group cannot satisfy our second criterion. In fact, no 2-design can be generated with $O(1)$ depth circuits over one- and two-qubit gates. We therefore choose to set $\mathbb{F} = \mathbb{C}_1^w$, where $ \mathbb{C}_1^w$ denotes the $w$-fold tensor product of the single-qubit Clifford group. 

Setting $\mathbb{F} = \mathbb{C}_1^w$ satisfies criteria (1), (2), and (3). It does not directly satisfy criterion (4), as $ \mathbb{C}_1^w$ does \emph{not} generate a 2-design. However, when combined with some simple and efficient data processing, $\mathbb{F} = \mathbb{C}_1^w$ does satisfy the fourth criterion. To understand why, observe that averaging over these fiducials performs a type of group-twirl \cite{gambetta2012characterization}. Fiducials from $\mathbb{C}_1^w$ implement the twirling map $\mathscr{T}$ that acts on $w$-qubit superoperators as
\begin{equation}
\mathscr{T}(\mathcal{E}) = \frac{1}{24^w} \sum_{L \in\mathbb{C}_1^{\otimes w}} \mathcal{U}(L) \mathcal{E} \mathcal{U}(L^{-1}).
\end{equation}
This twirl projects any superoperator onto the space spanned by $w$-fold tensor products of one-qubit depolarizing channels \cite{gambetta2012characterization} (in practice there will be errors in the fiducial subroutines, so they do not implement a perfect twirl. However,  it is known that the effect of twirling is robust under weak error  \cite{proctor2017randomized, wallman2017randomized, merkel2018randomized}). So averaging over the fiducials converts $\Lambda( \tilde{C}^{-1} Q C)$ into a stochastic Pauli channel with a distribution over Pauli errors that, for each of the $w$ qubits, has a uniform marginal distribution over the three Pauli errors, $X$, $Y$ and $Z$. This guarantees good (but not uniform) sensitivity to \emph{all} errors, because, although the $Z$ errors cause no observable failure --- \ie, the correct bit is output by the qubit on which the error occurs --- both $X$ and $Y$ errors flip the output bit of that qubit. So the rate of unobserved $Z$ errors can be inferred from the observed rate of bit flips.

This implies that there is a simple function of data that is equal to $F_e(\Lambda( \tilde{C}^{-1} Q C))$, up to contributions from errors in the initialization, readout and fiducial subroutines. For $k=0,1,\dots,w$, let $h_k$ denote the probability of the circuit producing a bit string whose Hamming distance from the target bit string is $k$ --- so $h_0 = S$ and $\sum_{k=0}^{w} h_k = 1$. For $k=0,1,\dots,w$, let $p_k$ denote the probability that $\mathscr{T}(\Lambda( \tilde{C}^{-1} Q C))$ induces any weight $k$ error --- so $p_0$ is the probability of no error, meaning that
\begin{equation}
p_0 = F_e(\Lambda( \tilde{C}^{-1} Q C)),
\end{equation}
and $\sum_{k=0}^{w} p_k = 1$. These distributions are related by
\begin{equation}
\vec{h} = M \vec{p},
\end{equation}
 where $M_{jk}$ is the probability that a weight-$k$ error causes $j$ bit flips on the target bit string. Now a weight-$k$ error causes $j$ bit flips if $j$ of the $k$ Pauli errors are not $Z$. Because the probability of all three Pauli errors is equal, this is simply given by
\begin{equation}
M_{jk} = {{k}\choose{j}} \frac{2^j}{3^k},
\end{equation}
for $j \leq k$, with $M_{jk} = 0$ for $j > k$. By inverting this equation we obtain:
\begin{equation}
p_0 = \sum_{k=0}^{w} \left( - \frac{1}{2}\right)^k h_k.\label{eq:p0}
\end{equation}

The Hamming distance distribution can be efficiently estimated from data (it is a distribution over $w+1$ elements). So we can use this relationship to efficiently estimate the process fidelity of $\Lambda( \tilde{C}^{-1} Q C)$ --- up to contributions from errors in the initialization, readout and fiducial subroutines, which, if desired, could be estimated and removed using standard techniques \cite{magesan2011scalable}. It would therefore be well-motivated to use the right-hand-side of Eq.~\eqref{eq:p0}, in place of $S$ or $P$ (the polarization), as a quantifier of how successfully a mirror circuit with randomized single-qubit Clifford fiducials has run. We do not do so in this work, however, for two reasons. First, $S$ and $P$ are arguably more intuitive. Second, using $p_0$ instead of $S$ or $P$ makes little difference to our results, and no difference to our scientific conclusions. One of the reasons for this is that $p_0 \approx S$ when most of the observed incorrect bit strings are a large Hamming distance from the target bit string. This will typically occur when $C$ is a wide and deep circuit containing many two-qubit gates (which will spread errors).

\begin{figure*}[t!]
\includegraphics[width=18cm]{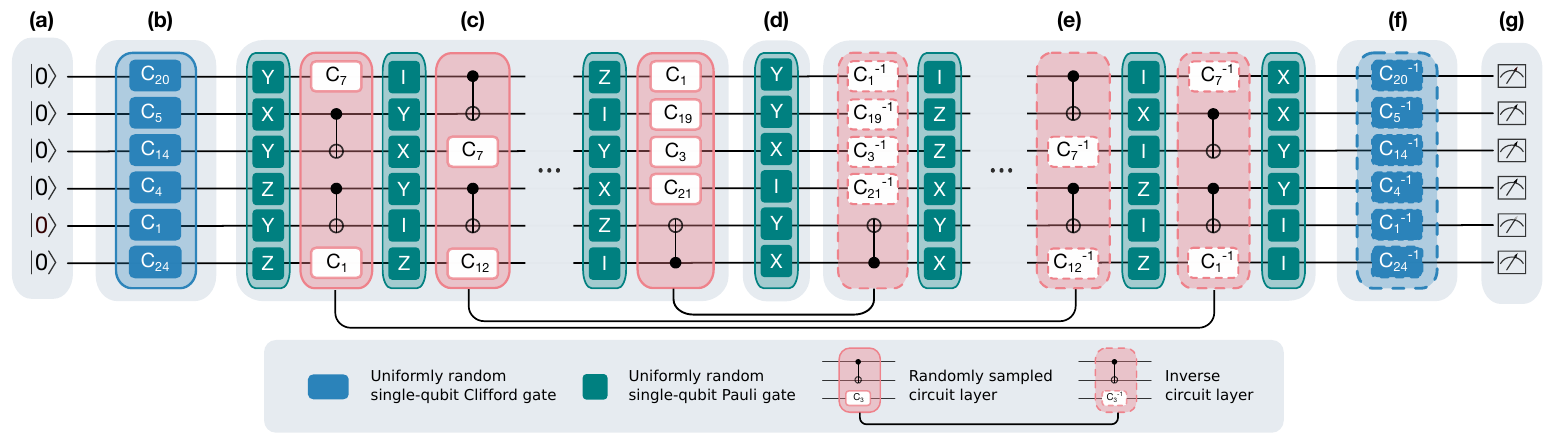}
\caption{\textbf{Randomized mirror circuits}. This figure shows a schematic of the randomized mirror circuits used in our experiments.  These circuits are sequences of $w$-qubit layers, of 5 distinct types:  initialization, random Pauli, random local Clifford, general Clifford, and readout.  Four of these layer types are completely standardized, but the set $\mathbb{L}_w$ of general Clifford layers can be configured to generate different benchmark ensembles, by specifying a distribution $\Omega$ over Clifford layers.  A similar approach was used in direct randomized benchmarking \cite{proctor2018direct}, and this construction can be generalized to non-Clifford circuits (see text). A width $w$ randomized mirror circuit with a \emph{benchmark depth} of $d$ consists of the following layers: \textbf{(a)} an initialization layer that prepares all $w$ qubits in $\ket{0}$; \textbf{(b)} a layer of uniformly random single-qubit Clifford gates on each qubit; \textbf{(c)} $\nicefrac{d}{4}$ independently sampled \emph{pairs} of layers, each comprising a layer of uniformly random Pauli gates followed by a layer sampled from $\Omega$; \textbf{(d)} a layer of uniformly random Pauli gates on each qubit; \textbf{(e)} the layers from step (c), but with their order reversed, each Pauli layer independently resampled, and each $\Omega$-random layer replaced with its inverse; \textbf{(f)} the inverse of the first layer of Clifford gates; and \textbf{(g)} a readout layer that measures each qubit in its computational basis. Randomized mirror circuits can have any width $w$ (here $w=6$) and any benchmark depth $d \geq 0$ that is an integer multiple of 4. Note that the full depth $d_0$ of the circuit is $d_0 = d + 5$ --- the benchmark depth ignores the five constant layers from steps (a, b, d, f, g). Benchmark depth is reported in the main text.}
\label{fig:rmcs}
\end{figure*}

\section{Randomized mirror circuits} \label{sec:rmcs}
Our experiments used two kinds of mirror circuits: \emph{randomized mirror circuits} and \emph{periodic mirror circuits}. In this appendix we define randomized mirror circuits.  Although the definitions in this appendix are self-contained, the mirroring transformations used to construct them were introduced and motivated in Appendix~\ref{sec:mirroring}.

\subsection{Definition}
Our experiments used randomized mirror circuits built from alternating layers of randomized Pauli gates and Clifford gates chosen from a sampling distribution $\Omega$ over a Clifford layer set $\mathbb{L}_w$.  In the second half of the circuit, each of the $\Omega$-random layers is inverted, but the Pauli layers are independently resampled randomly.  The sampling distribution $\Omega$ is configurable, and is used to vary and fine-tune the properties of the benchmark.  (A related construction plays a role in direct randomized benchmarking \cite{proctor2018direct}). 

A width-$w$ randomized mirror circuit with a benchmark depth of $d$ (see schematic in Fig.~\ref{fig:rmcs}) consists of: 
\begin{itemize}
\item[(a)] An initialization layer that prepares all $w$ qubits in $\ket{0}$.
\item[(b)] A layer of uniformly random single-qubit Clifford gates on each qubit. 
\item[(c)] A sequence of $\nicefrac{d}{4}$ independently sampled pairs of layers, where each pair consists of
\begin{enumerate}
\item A layer of uniformly random Pauli gates on each qubit.
\item A layer sampled from $\Omega$.
\end{enumerate}
\item[(d)] A layer of uniformly random Pauli gates on each qubit.
\item[(e)] The layers from step (c) in the reverse order with:
\begin{enumerate}
\item Each $\Omega$-random layer replaced with its inverse.
\item Each Pauli layer independently resampled.
\end{enumerate}
\item[(f)]The inverse of the Clifford layer from step (b).
\item[(g)] A readout layer that measures each qubit in the computational basis.
\end{itemize}
Randomized mirror circuits can have any width $w$, and any benchmark depth that is a multiple of four (\ie, $d=4k$ for some integer $k \geq 0$). As with all our benchmark circuits, note that the full depth $d_0$ of the circuit is $d_0 = d + 5$ --- the benchmark depth ignores the constant contribution of the five layers in steps (a), (b), (d), (f) and (g).

The bulk of a randomized mirror circuit is occupied by $\Omega$-random layers (which are the heart of the construction) and random Pauli layers.  The random Pauli layers play a simple role: they maximize the disorder of each circuit (no matter what $\Omega$ is used) and locally scramble errors.  They impose local basis randomization, which ensures that systematic, coherent errors on the layers almost surely do not not align or anti-align, and therefore do not interfere constructively or destructively over many circuit layers.  This has an effect somewhat similar to Pauli frame randomization \cite{knill2005quantum, wallman2015noise, ware2018experimental}. However, in contrast to Pauli frame randomization, the Pauli layers in our circuits are \emph{not} resampled each time the circuit is run. This is because our aim is not to convert all types of error into stochastic Pauli errors --- instead we are aiming to benchmark a processor's performance on disordered circuits. (Note, however, that a processor is free to implement our benchmarking circuits using randomized gate implementations. As discussed above, our construction is agnostic as to how the layers are implemented.)

The central random Pauli layer, which appears in \emph{all} our mirror circuits (including the periodic ones shown in Fig.~\ref{fig:pmcs}), plays a special role.  It prevents cancellation of errors between a circuit $C$ and the ``quasi-inverse'' circuit $\tilde{C}^{-1}$ used to mirror $C$. The alternating layers of randomized Pauli gates --- which only appear in \emph{randomized} mirror circuits --- play a similar role for each layer. They limit the degree to which coherent errors can systematically add or cancel between layers, on average. The $\Omega$-random layers also prevent systematic addition and cancelation, but the addition of the uniformly random Pauli layers causes the rate that coherent errors systematically add or cancel to only weakly depend on $\Omega$. This is convenient, because varying $\Omega$ is useful for generating varied and interesting benchmarking circuit ensembles.

The theory of direct randomized benchmarking \cite{proctor2018direct} can be used to show that the mean success probability of a randomized mirror circuit sampled according to $\Omega$ is closely related to the process fidelity of a $\Omega$-random circuit layer.  Similar relationships hold for other kinds of randomized circuit \cite{magesan2011scalable, boixo2018characterizing, cross2018validating}.  However, we do not use this relationship in this paper, so we do not pursue it further here.

\subsection{Circuit samplers}
Varying the distribution $\Omega$ over Clifford layers provides a way to tune and control important properties of the random mirror circuit benchmark.  One of the most important properties is the density of two-qubit gates within the circuits, which we denote $\xi$.  Each of our experiments used a distribution $\Omega$ over layers constructed from the following native gate set:
\begin{itemize}
    \item a set of single-qubit Clifford gates, $\mathbb{G}_1$, each of which may be applied to any qubit, and
    \item a single two-qubit Clifford gate that may be applied to any pair of connected qubits. 
\end{itemize}
We define $\Omega$ distributions constructively, by defining \emph{samplers} that generate layers.  Assigning two-qubit gates is the trickiest part of this sampling, and to do so we make use of an \emph{edge sampler} that we denote $\chi$. An edge sampler $\chi$ takes as input the connectivity graph of the $w$ qubits being benchmarked, and (usually) a parameter to control the number of edges that will be sampled.  It outputs a subset of edges that have no qubits (nodes) in common. This can be done in several ways, and we discuss the particular edge samplers we used in a moment. We can use this any such edge sampler $\chi$ to sample a $w$-qubit layer (which defines an $\Omega$) as follows:
\begin{enumerate}
\item Use $\chi$ to select a set of disjoint connected pairs of qubits from the $w$ available qubits.
\item Add a two-qubit gate on each edge selected in Step 1.
\item Assign a uniformly random single-qubit gate from $\mathbb{G}_1$ to each remaining qubit. 
\end{enumerate}
This sampling allows us to control the two-qubit gate density in the circuits, while guaranteeing that a typical circuit is always highly disordered. To specify a particular distribution $\Omega$, we only need to specify the edge sampler $\chi$. Below are the $\chi$ samplers used in our two experiments.

\subsubsection{The circuit sampling of experiment \#1}\label{sec:e1-sampler}
The randomized mirror circuits of experiment \#1 were sampled using a particularly simple edge sampler ($\chi_1$). It returns either \emph{zero} edges (with probability $\nicefrac{1}{2}$), or a single edge selected uniformly at random from the $w$-qubit connectivity sub-graph (with probability $\nicefrac{1}{2}$).  This sampling algorithm is not appropriate for arbitrarily large processors, because the expected two-qubit gate density of the circuits it generates ($\bar{\xi}$) goes to zero as $w \to \infty$. However, it is simple and transparent, and in the 1-16 qubit regime of our experiments it generates a useful array of circuits.  The low density of two-qubit gates ensures low enough error rates that we can actually probe how device performance varies with $d$ and $w$ (rather than seeing the success probability drop below measurable levels even for very small circuits).

\subsubsection{The circuit sampling of experiment \#2}\label{sec:e2-sampler}
The randomized mirror circuits of experiment \#2 are sampled using an edge sampler ($\chi_{\bar{\xi}}$) that we call the \emph{edge grab}. It is parameterized by the expected two-qubit gate density of the sampled circuits, $\bar{\xi}$. This sampler is designed for generating randomized mirror circuit benchmarks on arbitrarily large processors. 

Before we introduce the edge grab, we need to clarify our definition of two-qubit gate density ($\xi)$. The two-qubit gate density of a circuit $C$ with shape $(w,d)$ that contains $\alpha$ two-qubit gates is defined as $\xi = \nicefrac{2\alpha}{wd}$. If the circuit is thought of as a $w \times d$ lattice, this is the proportion of the lattice sites that are occupied by a two-qubit gate. In this work we use the benchmark depth to define $\xi$. 

The edge grab procedure $\chi_{\bar{\xi}}$ is defined as follows:
\begin{enumerate}
\item \emph{Select a candidate set of edges $E$}. Initialize $E$ to the empty set, and initialize $E_{r}$ to the set of all edges in the connected sub-graph of the $w$ qubits. Then, until $E_{r}$ is the empty set:
\begin{enumerate} 
\item[1.1] Select an edge $v$ uniformly at random from $E_{r}$.
\item[1.2] Add $v$ to $E$ and remove all edges that have a qubit in common with $v$ from $E_{r}$.
\end{enumerate}
\item \emph{Select a subset of the candidate edges}. For each edge in $E$, include it in the final edge set with a probability of $w\bar{\xi}/|E|$ where $|E|$ is the total number of edges in $E$. 
\end{enumerate}

The expected number of selected edges is $w \bar{\xi}$, so this sampler generates a $w$-qubit layer with an expected two-qubit gate density of $2\bar{\xi}$. Because only half of the layers in a randomized mirror circuit are sampled using $\chi$, randomized mirror circuits sampled according to the edge grab sampler have an expected two-qubit gate density of $\bar{\xi}$. Individual circuits' two-qubit gate density will fluctuate around this value, but the ensemble variance of $\xi$ converges to zero as the circuit size increases. This sampling algorithm has another nice property:  the probability of sampling a particular $w$-qubit layer $L$ is non-zero for every $L \in \mathbb{L}_w$ (except if $\bar{\xi}=0$ or $\bar{\xi} = \nicefrac{1}{2}$). The edge grab algorithm is invalid if $w\bar{\xi}/|E| > 1$ for any possible candidate edge set $E$. For an even number of fully-connected qubits, $\xi$ can take any value between 0 and $\nicefrac{1}{2}$ (note that $\nicefrac{1}{2}$ is the maximum possible $\xi$ in a randomized mirror circuit, as half the layers in these circuits contain only single-qubit gates). But for any other connectivity the maximum achievable value of $\bar{\xi}$ is smaller. In our experiments, we set $\bar{\xi} = \nicefrac{1}{8}$. This is an achievable value of $\bar{\xi}$ in the edge grab algorithm for all the processors that we benchmarked.

\begin{figure*}[t!]
\includegraphics[width=18cm]{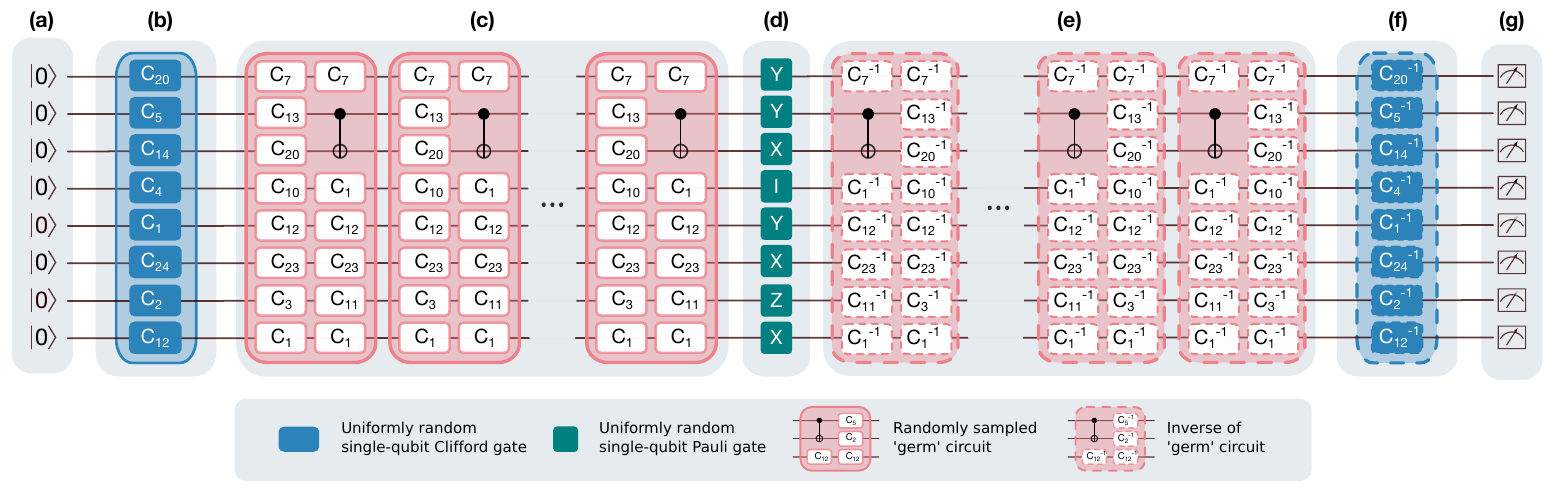}
\caption{\textbf{Periodic mirror circuits}. A schematic of the periodic mirror circuits that we use in our experiments. A width $w$ periodic mirror circuit with a benchmark depth of $d$ consists of the following layers: \textbf{(a)} initialization of all $w$ qubits in $\ket{0}$; \textbf{(b)} a layer of uniformly random single-qubit Clifford gates on each qubit; \textbf{(c)} $\nicefrac{d}{2d_g}$ repetitions of a `germ' circuit of depth $d_g$; \textbf{(d)} a layer of uniformly random Pauli operators on each qubit; \textbf{(e)} the layers from step (c) in the reverse order and with each layer replaced with its inverse; \textbf{(f)} the inverse of the first layer of Clifford gates; \textbf{(g)} readout of each qubit in the computational basis. If $\nicefrac{d}{2d_g}$ is not an integer then only some of the layers of the germ circuit are run in the last repetition of the germ in (c). In our experiments, the germ circuit is sampled at random, using an algorithm (detailed in the text) that generates a germ circuit with a two-qubit gate density $\xi \leq \nicefrac{1}{8}$. In the example shown here, the two-qubit density is exactly $\xi=\nicefrac{1}{8}$. Periodic mirror circuits can have any width $w$ ($w=8$ in this example) and any even benchmark depth $d \geq 0$. Note that the full depth $d_0$ of the circuit is $d_0 = d + 5$. As with all our mirror circuits, we have removed the constant contribution of the five layers in (a), (b), (d), (f) and (g) from our definition of the benchmark depth. It is the benchmark depth that is reported in the main text.}
\label{fig:pmcs}
\end{figure*}

\section{Periodic mirror circuits} \label{sec:pmcs}
Our experiments consisted of running two types of mirror circuit benchmark: \emph{randomized mirror circuits} and \emph{periodic mirror circuits}. In this appendix we define the class of periodic mirror circuits, and the specific periodic mirror circuits that we ran in our experiments. These circuits are constructed using the mirroring circuit transformations introduced in Appendix~\ref{sec:mirroring}, but note that the definitions in this appendix are self-contained.

\subsection{Definition}
Fig.~\ref{fig:pmcs} illustrates the form of our periodic mirror circuits. They are based on repetitions of a low-depth \emph{germ circuit} $C_g$, named following the terminology of gate set tomography \cite{blume2016certifying}. For a given germ circuit $C_g$ of shape $(w,d_g)$, a width-$w$ periodic mirror circuit with a benchmark depth of $d$ consists of:
\begin{itemize}
\item[(a)] An initialization layer placing all $w$ qubits in the $\ket{0}$ state.
\item[(b)] A layer of uniformly random single-qubit Clifford gates on each qubit. 
\item[(c)] A depth $\nicefrac{d}{2}$ circuit constructed by repeating $C_g$ $\left\lceil \nicefrac{d}{2d_g} \right\rceil$ times, and removing the final $(\nicefrac{d}{2} \;\mathrm{mod}\; d_g)$ layers. 
\item[(d)] A layer of uniformly random Pauli operators on each qubit.
\item[(e)] The layers from step (c) in the reverse order and with each layer replaced with its inverse.
\item[(f)] The inverse of the first layer of Clifford gates in step (b).
\item[(g)] A layer reading out each qubit in the computational basis.
\end{itemize}
Periodic mirror circuits can have any width $w$, and any benchmark depth $d$ that is an integer multiple of two. As with all our benchmark circuits, note that the full depth $d_0$ of the circuit is $d_0 = d + 5$, so we have removed the constant contribution of the five layers in (a), (b), (d), (f) and (g) from our definition of the benchmark depth.

\subsection{Selecting the germ circuit}
Defining a specific set of periodic mirror circuits --- or a specific distribution over periodic mirror circuits --- requires choosing a method for selecting germ circuits $C_g$. Repeating a specific germ amplifies the effect of some errors, while suppressing others \cite{blume2016certifying}. For example, a single-qubit germ circuit consisting of a single $X$ gate amplifies coherent over/under-rotation errors in the $X$ gate, but it suppresses the effect of ``tilt'' errors --- \ie, $Y$ or $Z$ Hamiltonians that change the rotation axis of the $X$ gate. (For example, if $X$ is implemented perfectly except that it is followed by an erogenous small $\hat{z}$-axis coherent error, then a circuit consisting of an even number of $X$ gates composes to an exact identity). It is possible, in principle, to construct germs that, collectively, amplify all the parameters in a specific error model \cite{blume2016certifying}. But a general model of Markovian errors on $w$ qubits contains $16^w-4^w$ parameters \emph{per layer}, and amplifying all those parameters is infeasible. We could define a much smaller error model and construct germs that amplify all its parameters, but this is only well-motivated if that smaller model accurately describes the tested processors. We therefore take a different approach: we sample germs at random, using an algorithm that is biased towards amplifying parameters that are likely to be physically important.
 
\subsubsection{The germ selection of experiment \#2}\label{sec-germ-e2}
We ran periodic mirror circuits in experiment \#2. Here we describe the germ sampling algorithm that we used. It is composed of two steps. The first step constructs a germ circuit composed of only single-qubit gates, and the second step replaces some of these gates with two-qubit gates. This protocol is somewhat complicated, but was designed for investigating our specific scientific question --- the effect of circuit order on circuit failure rates --- and it is not intended as a general-purpose germ selection routine. We expect that different algorithms for generating periodic mirror circuits will be useful for, \eg, creating standardized benchmarks.

\vspace{0.2cm}
\noindent \textbf{Step 1}: The first step in our algorithm is to create a width-$w$ germ circuit $C_g$ that contains only single-qubit gates from some set $\mathbb{G}_1$ (in our experiments, $\mathbb{G}_1$ was the 24-element group of single-qubit Clifford gates). Our specific sampling algorithm was:
\begin{enumerate}
\item \emph{Select a germ depth $d_{g}$}. We do this by setting $d_{g} = 2^x$ with probability $1/2^{x+1}$ for $x=0,1,2,\dots$, and then truncating the depth to $8$. That is, if $d_{\rm g} > 8$ set $d_{\rm g} = 8$. An exponentially decaying probability truncated at depth 8 is useful because a depth-$d$ circuit constructed by repeating a germ of length $d_g$ is only periodic if $d > d_g$. Current processors cannot run very deep circuits without an error almost certainly occurring, so we can only study periodicity by repeating relatively shallow germ circuits.
\item \emph{Select a local germ for each qubit}. For each of the $w$ qubits, indexed by $i$, we independently select a local germ $C_{l,i}$ by:
\begin{enumerate}
\item[2.1] Setting $d_{l,i} = 2^x$ with probability $1/2^{x+1}$ (for $x=0,1,2,\dots$), and, if the selected $d_{l,i}$ is greater than $d_{g}$, then setting $d_{l,i}  = d_{g}$.
\item[2.2] Setting $C_l$ to a uniformly random depth-$d_{l}$ sequence of single-qubit gates from $\mathbb{G}_1$. 
\end{enumerate}
\item \emph{Combine the local germs}. Construct a germ circuit $C_g$ of depth $d_g$ by combining the $w$ independently selected local germs in parallel. To create a depth $d_g$ circuit, the local germ for qubit $q$ is repeated $\nicefrac{d_g}{d_{l,q}}$ times, where $d_{l,q}$ is the depth of that local germ. By construction, this consists of an integer number of repetitions of each local germ.
\end{enumerate}
We designed a sampling algorithm that has a strong bias towards shallow local germs --- \eg, the marginal probability of a depth 1 local germ is $\nicefrac{3}{4}$ --- because depth 1 germs amplify a particularly important class of errors that includes coherent over/under-rotations.

\vspace{0.2cm}
\noindent \textbf{Step 2}: The second step in our randomized germ selection algorithm is to replace some of the gates in $C_g$ with two-qubit gates (unless $w=1$, in which case this step is skipped). In order to test the hypothesis that periodic circuits perform worse than disordered circuits, we chose an algorithm that generates germs with a two-qubit density of $\xi \leq \nicefrac{1}{8}$, because $\nicefrac{1}{8}$ is the expected two-qubit gate density in the randomized mirror circuits that we ran alongside these periodic mirror circuits (see above, and note that these experiments are detailed further in Appendix~\ref{sec:b2}). This then means that, if we observe worse performance on periodic mirror circuits, this cannot be explained by higher $\xi$. The algorithm that we used, defined for $w>1$, is as follows:
\begin{enumerate}
\item Set $r$ to the minimum positive integer that satisfies $\nicefrac{2}{r d_g w} < \nicefrac{1}{8}$, where $d_g$ is the current germ's depth, and then replace the germ circuit $C_g$ with $r$ repetitions of $C_g$. This means that we can place at least one two-qubit gate within the germ and still obtain $\xi \leq \nicefrac{1}{8}$.
\item For each layer in the updated germ select a set of edges $E_{l}$, with $l= 1, 2, \dots, r d_g$, using the first step of the ``edge grab'' sampling algorithm (see Appendix~\ref{sec:rmcs}). Then combine them into a single set $E_g $ consisting of layer-index and edge pairs.
\item Place $n = \nicefrac{r d_g w}{16}$ two-qubit gates into the germ, by
\begin{enumerate}
\item selecting $n$ layer-index and edge pairs from $E_g$, uniformly at random, and
\item replacing the one-qubit gates at each of these positions in the germ with a two-qubit gate.
\end{enumerate}
\end{enumerate}

Note that germs generated via this algorithm have a two-qubit gate density of $\xi \leq \nicefrac{1}{8}$. However, periodic mirror circuits generated from these germs \emph{can} have a two-qubit gate slightly density above $\xi$, because a germ is only partially repeated in a depth $d$ periodic mirror circuit if the germ circuit's depth is not a factor of $\nicefrac{d}{2}$.

\section{Predicting mirror benchmarks from a processor's error rates} \label{sec:predictions}
Figure 2c of the main text compares the \emph{measured} results of our benchmarks with the \emph{predicted} performance based on the published error rates provided for each of the quantum processors. In this appendix we explain how we obtain these predictions. 

The set of ``error rates'' $\{\epsilon\}$ provided for a given processor can consist of many different performance metrics estimated in many different ways. For example, the entire error rate set $\{\epsilon\}$ could consist of a single heuristic error rate for the entire processor. At the opposite extreme, $\{\epsilon\}$ could consist of all the parameters of a detailed process matrix error model fit using, \eg, gate set tomography \cite{blume2016certifying}. For the processors in our experiments, the contents of the error rate sets lie between these two extremes. They include summary error rates for the native logic operations, with the gate error rates measured by randomized benchmarking. The error rate set represents a valuable description of a processor's performance, but it does not immediately imply a detailed predictive model for the processor. In order to predict a circuit's success probability from the provided error rates $\{\epsilon\}$, we need to construct a predictive formula or model in which the only parameters are (1) these error rates, and (2) the circuit.

\subsection{Standard error rates}\label{sec:standard-error-rates}
The exact metrics that constitute the reported error rates, $\{\epsilon\}$, display some minor variation across processors. But all of these can be straightforwardly transformed into a ``standard form'' capable of describing each of the processors we benchmarked. This standard form consists of:
\begin{itemize}
\item The estimated entanglement infidelity for each available single-qubit gate $G$ on each possible target physical qubit $i$, which we denote $\epsilon(G_i)$.
\item The estimated entanglement infidelity for each available two-qubit gate, indexed by the target physical qubits, $i$ and $j$, which we denote $\epsilon(G_{i,j})$. 
\item A readout error rate $\epsilon(i)$ for each physical qubit $i$ defined by
\begin{equation}
\epsilon(i) =\frac{1}{2}\big( \;\mathsf{Pr}(1 \vert 0) + \mathsf{Pr}(0 \vert 1)\;\big),
\end{equation}
 where $\mathsf{Pr}(x\vert y)$ is the probability of reading out $x$ on qubit $i$ after preparing that qubit in the state $\ket{y}$.
\end{itemize}
Initialization errors are not reported separately, and are instead implicitly included in the readout error rate, so $\epsilon(i)$ can be thought of as an average state preparation and measurement (SPAM) error. Further note that this standard form explicitly utilizes the \emph{entanglement} infidelity, rather than the \emph{average gate} infidelity that is the usual error metric associated with randomized benchmarking \cite{magesan2011scalable} (and which is the error metric used by IBM Q and Rigetti). The two infidelities are simply related to each other by the linear rescaling of Eq.~\eqref{eq:Fe-Fa}.

\subsection{Constructing a predictive model}
This standard form given above for the error rate set $\{\epsilon\}$ does not directly constitute a predictive model. Below we describe several increasingly detailed approaches for converting the descriptive error rates into predictive models of circuit success probabilities, and we highlight the method that we actually used.

\subsubsection{A simple error accumulation formula}
The simplest approach to predicting the success probability $S$ of a circuit $C$ is to compute the probability that \emph{no} error happens over the course of the circuit. This is simply the product of one minus the error rates of all the operations in $C$. In the small error and small circuit limit, the predicted \emph{failure} rate $(1-S)$ is then approximately the sum of the error rates of the constituent operations. So for the circuit $C=RL_dL_{d-1} \cdots L_2L_1 I$, we have:
\begin{equation}
S = s(R)s(L_d) \cdots s(L_2)s(L_1) s(I), \label{eq:S-naive-pred}
\end{equation}
where $s(L)$ is the success probability of layer $L$ given by the product of one minus the error rates of the layer's constituent operations:
\begin{itemize}
\item For the initialization layer $I$, $s(I)= 1$. As discussed above, errors in the initialization are captured by the ``readout'' error rates.
\item For a gate layer $L$ 
 \begin{align}
s(L) &=  \prod_{G \in L} (1-\epsilon(G)),
\end{align}
where the product is over the particular one- and two-qubit gates (on particular qubits) from which $L$ is constructed. 
\item  For the readout layer $R$
 \begin{align}
s(R) &=  \prod_{i \in \mathbb{Q}} (1-\epsilon(i)), \label{eq:readout-layer-error}
\end{align}
where $\mathbb{Q}$ is the set of indices of the qubits on which $C$ acts.
\end{itemize}

\subsubsection{The global depolarization model}
Equation~\eqref{eq:S-naive-pred} is simple and intuitive, but it is flawed. This is because it implicitly assumes that two or more errors cannot cancel, and so it predicts that $S\to 0$ as circuit depth $d \to \infty$ rather than $S \to \nicefrac{1}{2^w}$. So, instead, we use a formula that corrects for this. We predict $S$ using

\begin{equation}
S = \nicefrac{1}{2^w} + (s(R) - \nicefrac{1}{2^w})\lambda(L_d)\lambda(L_{d-1}) \cdots \lambda(L_1),
\end{equation}
where 
\begin{equation}
\lambda(L) =  \frac{1}{1-4^w}\left( 1 -4^w \prod_{G \in L} \left( 1-\epsilon(G) \right) \right).
\end{equation}
Although this formula might seem much more complex than Eq.~\eqref{eq:S-naive-pred}, it follows simply from modeling the error in each gate layer as a \emph{global} $w$-qubit depolarizing channel [see Eq.~\eqref{eq:global-dep}] with an entanglement fidelity equal to the product of the entanglement fidelities of the constituent gates --- which is how entanglement fidelity composes under tensor products. Moreover, note that this formula for $S$ depends \emph{approximately} only on the number of times each gate (and readout) appears in the circuit. This holds only approximately because errors compose differently when they occur in parallel or in serial (errors on different qubits that occur in the same layer cannot cancel, where errors on different layers \emph{can} cancel). 

\subsubsection{The local depolarization model}
An alternative model in which to embed the error rates is a local depolarizing model. In this model, each one-qubit gate $G_i$ is modeled as the perfect unitary followed by the one-qubit depolarizing channel $\mathcal{D}_{1,\epsilon(G_i)}$,  and each two-qubit gate $G_{i,j}$ is modeled as the perfect unitary followed by the two-qubit depolarizing channel $\mathcal{D}_{2,\epsilon(G_{i,j})}$ [again, see Eq.~\eqref{eq:global-dep} for the definition of a $w$-qubit depolarizing channel]. This is arguably more physically well-motivated than the global depolarizing model, because it is consistent with the characterization experiments from which the errors rates are extracted --- that is, under this model, one- and two-qubit randomized benchmarking will return the error rates used in the model (up to scaling differences between average gate and entanglement infidelity).  However, unlike the previous two models, the local depolarization model does not lend itself to a compact, analytical formula for the success probability. In order to make predictions from a local depolarizing model it is necessary to simulate the circuit. 

Because our benchmarks use Clifford circuits, weak simulation (\ie, sampling from the circuit's output distribution) under local depolarization is efficient in both circuit depth $d$ and width $w$. Strong simulation (\ie, computing the success probability exactly) is expensive, however, scaling exponentially in $d$. Somewhat surprisingly, the success probabilities predicted by this model are typically \emph{approximately} the same as those predicted by a corresponding global depolarizing model. This is because, under either model, a circuit's success probability is controlled only by (1) the rate that errors occur and (2) the rate that errors cancel. The error occurrence rate is equal in both models, and the error cancellation rate is \emph{almost} equal in both models \emph{unless} there is a large variance in the gate error rates on different qubits. For these reasons, we choose to use the global depolarizing model in this work.

\section{Experiment \#1} \label{sec:b1}
The purpose of this appendix is to describe the benchmarking experiments and data analysis summarized in Fig.~1d of the main text. Throughout these appendices we refer to these benchmarking experiments collectively as \emph{experiment \#1}. This appendix is not intended to be self-contained, and we make explicit references to earlier appendices when necessary. This appendix consists of two parts: in Appendix~\ref{sec:e1-exps} we detail the experiments, and in Appendix~\ref{sec:e1-analysis} we detail the data analysis.

\subsection{Experimental details}\label{sec:e1-exps}
Experiment \#1 used randomized mirror circuits to benchmark each of the twelve processors shown schematically in Fig.~1d. The benchmarking circuits were designed using a procedure we refer to as \emph{benchmark \#1} that can be applied to any gate-model quantum information processor. Benchmark \#1 has two notable properties: First, it was designed specifically for processors with fewer than $\sim 20$ qubits (in contrast to the benchmark of experiment \#2, described in Appendix~\ref{sec:b2}). Second, these benchmarking experiments took place over a period of time during which our methods were still under active development (the experiment dates range from July 2018 to November 2019), and so some minor aspects of the procedure changed over this time. We will note these changes explicitly as we introduce the benchmark. It was not possible to re-run all of the experiments with identical procedures, because not all of the processors were available for the full period of this research (in particular, IBM Q Rueschlikon, IBM Q Tenerife, Rigetti Agave and Rigetti Aspen-6 were no longer available in autumn 2019). This contrasts with experiment \#2 (see Appendix~\ref{sec:b2}) which is entirely standardized across the eight tested processors.

\subsubsection{Circuit benchmarking algorithms}\label{sec:b-alg}
Benchmark \#1  is an algorithmic approach for generating mirror circuit benchmarks to run on generic gate-model quantum information processors. It utilizes the following processor-specific information:
\begin{enumerate}
    \item A single-qubit gate set $\mathbb{G}_1$. We assume that all gates in $\mathbb{G}_1$ can be applied to any qubit on the processor. 
    \item A two-qubit gate $G_2$. Without loss of generality, this gate is assumed to be asymmetric and may be applied to any adjacent qubits on the processor's directed connectivity graph. (Fig.~1d displays the undirected connectivity graphs for each of the twelve processors we tested).
\end{enumerate}
The algorithm then generates a suite of circuits to run on the target processor. The qubits in each circuit are explicitly assigned to specific physical qubits, and the circuits are composed of layers built from $\mathbb{G}_1$ and $G_2$ gates allowed by the device's connectivity. The motivation for choosing this sort of benchmarking circuits is covered in detail in Appendix~\ref{sec:layer-set}.

\subsubsection{Gate set}\label{sec:b1-gates}
The IBM Q and Rigetti processors use different native gate sets. For this reason, we chose different gate sets for IBM Q and Rigetti processors:
\begin{itemize}
\item IBM Q processors:
    \subitem $\mathbb{G}_1 = \mathbb{C}_1$, where $\mathbb{C}_1$ is the set of all 24 single-qubit Clifford gates,
    \subitem $G_2 = \cnot$.
\item Rigetti processors:
    \subitem $\mathbb{G}_1$ comprises an idle gate, the three other single-qubit Pauli gates, and $\pm \nicefrac{\pi}{2}$ rotations around $\hat{x}$ and $\hat{z}$. $\mathbb{G}_1$ is a strict subset of $\mathbb{C}_1$.
    \subitem $G_2 = \cphase$. 
\end{itemize}
In contrast, in experiment \#2 we standarized the single-qubit gate set (to $\mathbb{G}_1 = \mathbb{C}_1$).
\subsubsection{Circuit shapes}
The first step in the benchmark \#1 algorithm is to select the set of circuit shapes at which to construct benchmarking circuits. For an $n$-qubit processor, we chose the circuit shapes $(w,d)\in\mathbb{W}_n\times\mathbb{D}$, where:
\begin{align}
    \mathbb{W}_n &= \{\,2^j \;\vert\; j \in [0\,..\,\lfloor \log_2(n)\rfloor\,] \,\}  \;\cup\; \{n\}\\
        &=\{1,2,4,\ldots,n\} \\
    \mathbb{D} &= \{ \,4 \lfloor  1.4^j \rfloor \; \vert\; j \in [1..13] \}\\
        &= \{ 0, 4, 8, 12, 20, 28, 40, 56, 80, 112, 160, 224, 316 \}
\end{align}
The circuit widths $w\in\mathbb{W}_n$ are powers of 2, with $w=n$ additionally included as the largest width (regardless of whether $n$ itself is a power of 2 or not). The benchmark depths $d\in\mathbb{D}$ are approximately exponentially spaced. We enforce that all depths are an integer multiple of 4, as this is a requirement of randomized mirror circuits (see Appendix~\ref{sec:rmcs}). For six of the twelve experiments (IBM Q Rueschlikon, IBM Q Melbourne, IBM Q Tenerife, Rigetti Agave, Rigetti Aspen 4 and Rigetti Aspen 6) we excluded depths $\{56, 112, 160, 224, 316\}$. For the other six experiments, which were all on 5-qubit IBM Q processors, we iteratively excluded the largest depth as the width was increased, \ie, shapes $(2,316)$, $(3,316)$, $(3,224)$, $(5,316)$, $(5,224)$, and $(5,160)$ where excluded. These choices were made in order to reduce the number of circuits required and/or due to limitations in what a particular processor could run. 

\subsubsection{Circuit embeddings}
For any circuit width $w < n$ we must select a set (or several sets) of $w$ connected physical qubits \emph{before} generating our benchmark circuits. This is because our benchmarks use layers of native gates, so we need to ensure that our circuits respect the connectivity constraints of the $w$ selected physical qubits. For most common connectivity graphs, as $n$ increases there is a rapidly increasing number of distinct connected sets of $w$ qubits for any non-extremal width (\ie, a width $w$ satisfying $1\ll w\ll n$).  For each width $w$ we select multiple width-$w$ sets $s_w$. We do so as follows:
\begin{itemize}
\item For a processor of $n \leq 5$ qubits, for each width we select each possible set of $w$ connected qubits. 
\item For a processor of $n > 5$ qubits, for each width $w$ we select $\lceil \nicefrac{n}{w} \rceil$ sets of $w$ connected qubits whereby every qubit is in at least one set of each size (here  $\lceil \cdot \rceil$ denotes the ceiling function, \ie, rounding up).
\end{itemize}

\subsubsection{Circuit sampling}
For each processor, each circuit shape $(w,d)$, and each chosen set of $w$ qubits ($s_w$) we sampled 40 shape-$(w,d)$ randomized mirror circuits acting on those $w$ qubits. These circuits were constructed using the $\chi_1$ sampler introduced in Appendix~\ref{sec:e1-sampler}. The code that we used to perform this sampling has been incorporated into the open-source software package \texttt{pyGSTi} \cite{nielsen2020probing, pygstiversion0.9.9.1}.

\subsubsection{Experimental details}
We ran benchmark \#1 on the twelve processors shown in Fig.~1d. The experiments were run using the online access services of IBM Q \cite{ibmq2} and Rigetti \cite{rigetti-qcs}. Both IBM Q and Rigetti routinely recalibrate their processors; all of the circuits were run within a single calibration window. Each circuit was repeated 1024 times, except for the experiments on Rigetti Agave, where each circuit was repeated 1000 times. For our first six experiments (IBM Q Melbourne, IBM Q Rueschlikon, IBM Q Tenerife, Rigetti Agave, Rigetti Aspen 4, and Rigetti Aspen 6), equal-depth, single-qubit circuits on different qubits were implemented simultaneously \cite{gambetta2012characterization}. That is, for each processor and each circuit depth $d$, the $40n$ width-1 circuits were combined into 40 width-$n$ circuits consisting of running one of the 40 depth-$d$ circuits for each qubit in parallel. For the circuit embedding strategy for processors of more than five qubits, this approximately halves the total number of circuits that need to be run. In an ideal processor, running these circuits in parallel has no effect on their outputs, but for real processors this is typically not the case, due to pulse spillover and other crosstalk effects \cite{gambetta2012characterization, sarovar2019detecting}. So, in our later six experiments (IBM Q Yorktown, IBM Q Ourense, IBM Q Essex, IBM Q London, IBM Q Vigo, and IBM Q Burlington), we ran the width-1 circuits separately. We did not parallelize any of the $w>1$ circuits in any of our experiments, as two-qubit gate crosstalk is known to often be a significant effect in superconducting chips \cite{rudinger2018probing, proctor2018direct, harper2019efficient}.

\subsection{Data analysis}\label{sec:e1-analysis}
The results of experiment \#1 are summarized in the \emph{volumetric benchmarking} plots \cite{blume2019volumetric} of Fig.~1d. Here we explain the data analysis used to generate these plots. In this appendix we use notation that distinguishes between a circuit's true success probability ($S$) and an observed success probability ($\hat{S}$) obtained from a finite number of repetitions of that circuit. 
As noted above, for some processors we left out depth 56 in order to reduce the total number of circuits. For these processors, in the plots in Fig.~1d the boxes (and frontiers) at depths 40 and 80 are stretched horizontally to meet at depth 56, so that there is no empty space in the plots.
 
\subsubsection{Circuit polarization}\label{sec:pol}
For each circuit that we ran, we calculate the observed polarization
\begin{equation}
\hat{P} = (\hat{S} -  \nicefrac{1}{2^w})/(1- \nicefrac{1}{2^w}),
\end{equation}
where $\hat{S}$ is that circuit's observed success probability and $w$ is the circuit's width. The polarization removes few-qubit effects.  If a processor is subject only to depolarizing noise, then $1 \geq S  \geq \nicefrac{1}{2^w}$, and as $d \to \infty$ then $S \to \nicefrac{1}{2^w}$ for any shape $(w,d)$ circuit. This is because a deep circuit will output $w$ uniformly random bits. So, under this noise model, $P\to 0$ as $d\to \infty$ and $1 \geq P \geq 0$ for any width circuit. Note that $\hat{P}$ can be negative, which can be caused by finite sampling or because $P$ itself can be negative under more general error models.

\subsubsection{Selecting the best qubits}
For each benchmarked circuit shape $(w,d)$ and each benchmarked subset of $w$ qubits ($s_w$) we ran 40 distinct randomized mirror circuits and measured 40 corresponding observed polarizations $\hat{P}$. These polarizations are collected into a set $\hat{\mathbf{P}}(w,d,s_w)$ for each circuit shape and qubit set. For each width $w$, the first step in our analysis is to identify the single set of qubits $b_{w}$ that we deem to have performed the best on our benchmarking circuits. We then discard the data for all other qubit sets, and generate volumetric benchmarking plots using only $\hat{\mathbf{P}}(w,d) \equiv \hat{\mathbf{P}}(w,d, b_w)$.

We select $b_w$ to be the $w$ qubits with the largest $d_{\text{mean}}$, where $d_{\text{mean}}$ is the smallest depth at which the mean polarization drops below $\nicefrac{1}{e}$. When more than one of the benchmarked sets of $w$ qubits have the same value for $d_{\text{mean}}$ we choose the set of qubits with the largest mean polarization at that depth. This process means that we have selected the $w$ qubit subsets that maximize the depth of the processor's mean polarization $\nicefrac{1}{e}$ frontier, which is the solid black line in each panel of Fig.~1d (discussed below). 

\subsubsection{Maximum, minimum and mean polarization}
In the volumetric benchmarking plot for each processor in Fig.~1d, we display the best, average, and worst case polarization versus circuit shape for the best-performing sets of qubits. That is, at each circuit shape $(w,d)$, we plot the maximum, mean, and minimum of $\hat{\mathbf{P}}(w,d)$. A circuit's polarization can be negative, so we truncate each of our performance metrics to zero. In the case of the mean, this truncation occurs \emph{after} averaging. 

\subsubsection{Performance frontiers}
In each panel of Fig.~1d we plot three frontiers, corresponding to the circuit shapes at which the maximum, mean and minimum polarizations drop below $\nicefrac{1}{e}$. For performance frontiers it is often useful to account for finite sampling effects, \ie, to adjust for the finite number of repetitions of each circuit. Details of this statistical analysis are given below. For now, we assume a statistic-specific function $f$ that takes $\hat{\mathbf{P}}(w,d)$, the set of observed polarizations, and returns ``pass'' or ``fail'' for that circuit shape.  

It is convenient to enforce that the frontier be monotonic, in the sense that as width or depth is increased the boundary is guaranteed to only be crossed once. So, given an $f$ function, the frontier is calculated as follows:
\begin{enumerate}
\item For each tested circuit shape $(w, d)$ use $f(\hat{\mathbf{P}}(w,d))$ to designate that circuit shape as  a ``pass'' or a ``fail''.
\item Set the frontier to the border of the largest region $R$ for which, if $(w^*,d^*) \in R$, $w\le w^*$, and $d\le d^*$, then $f(\hat{\mathbf{P}}(w,d))=$ ``pass''.  
\end{enumerate}

Of course, frontiers may be calculated for any threshold value. We choose $\nicefrac{1}{e}$ because circuit polarization will decay exponentially with the benchmark depth under the simplest error model --- uniform, layer-independent depolarization (\ie, a global depolarizing channel with the same error rate for every circuit layer). When the decay is approximately exponential the frontier is a visual representation of the rate of this approximately exponential decay. 

\subsubsection{Accounting for finite sample fluctuations}\label{sec:hypothesis}
The most appropriate method for accounting for the finite number of repetitions of each circuit ($N$) when calculating a statistic's frontier depends on the inferences that will be made from that frontier. In the case of the mean, we use the ``raw'' frontier that has no finite sampling adjustments. That is, for the mean, we use an $f$ function that simply returns ``pass'' if the mean of $\hat{\mathbf{P}}(w,d)$ is above $\nicefrac{1}{e}$ and otherwise it returns ``fail''. This is an unbiased estimate of whether the mean polarization is above or below the threshold value, and so it is a natural choice. 

Different choices are possible of course, and \emph{statistical hypothesis testing} \cite{lehmann2006testing} provides a rigorous framework for constructing broad classes of thresholding functions. For the case of the mean, for instance, one may desire a function $f$ that hypothesizes the mean is above a threshold, returning  ``fail'' if and only if there is \emph{statistically significant} evidence that the mean is below the $\nicefrac{1}{e}$ threshold value. For the maximum and minimum frontiers, we will utilize this hypothesis testing framework exclusively. 

In the main text we use the observation of a substantial discrepancy between the maximum and minimum frontiers as evidence that that processor is subject to highly structured errors. We therefore chose to calculate the maximum and minimum frontiers using a statistical hypothesis test that is designed so that the boundaries will be equal if there is no statistically significant evidence in the data to the contrary. This therefore guarantees that any observed discrepancy is not simply an artifact of finite $N$.

At each circuit shape, we start from the null hypotheses $H_0$ that either $H_{\uparrow}$ is true or $H_{\downarrow}$ is true, where:
\begin{itemize}
\item $H_{\uparrow}$ is the hypothesis that every circuit of this shape that we ran has a polarization that is above the $\nicefrac{1}{e}$ threshold.
\item $H_{\downarrow}$ is the hypothesis that every circuit of this shape that we ran has a polarization that is below the $\nicefrac{1}{e}$ threshold. 
\end{itemize}
Note that these hypotheses are about the circuits that we ran, \emph{not} the distribution of circuits from which they were sampled. Starting from the $H_0$ hypothesis at each circuit shape encodes our aim of starting from the assumption that the maximum and minimum frontiers are equal. Only if we can reject $H_0$ at a given circuit shape, using a statistical hypothesis test with 5\% significance, do we assign ``pass'' to the maximum polarization and ``fail'' to the minimum polarization. Otherwise we assign ``pass'' or we assign ``fail'' to both statistics (using the strategy outlined below). 

To test the null hypothesis $H_0$ at a given circuit shape, we perform two statistical hypothesis tests at $5\%$ significance --- one that tests for evidence to reject $H_{\uparrow}$, and one that tests for evidence to reject $H_{\downarrow}$. Because we must reject both $H_{\uparrow}$ and $H_{\downarrow}$ to reject $H_0$, the significance of this type of test for $H_0$ is $5\%$. The two tests that we use are equivalent, so we only describe the test of $H_{\uparrow}$. This test is more simply described in terms of each circuit's observed success probability $\hat{S}$, rather than in terms of the polarizations.

The statistical hypothesis test of $H_{\uparrow}$ that we use consists of $K$ log-likelihood ratio tests \cite{rudinger2018probing}, where $K$ is the number of circuits of that shape (here $K=40$). We test whether each observed success probabilities $\hat{S}$ is consistent with the null hypothesis that it is the average of $N$ draws from a 0/1-valued ``coin'' with a probability $S$ to output 1 that is above $T_S = (1 + \nicefrac{1}{2^w})\nicefrac{1}{e} + \nicefrac{1}{2^w}$ ($T_S$ is the $\nicefrac{1}{e}$ polarization threshold rescaled to the equivalent success probability threshold). There are $K$ hypothesis tests performed, and so to maintain the test significance to $5\%$ we must account for this. We do so using the Benjamini-Hochberg procedure \cite{benjamini1995controlling}. We then reject $H_{\uparrow}$ if any of the tests indicate that their circuit's $S$ is below $T_S$. This is similar to rejecting $H_{\uparrow}$ if the smallest p-value in these $K$ tests is smaller than $\nicefrac{0.05}{K}$, which is the well-known Bonferroni correction, but this testing procedure is more powerful. (The Benjamini-Hochberg procedure with $\alpha$ significance guarantees that, if all the tested null hypothesis are true, the probability of rejecting one or more null hypotheses is at most $\alpha$. This is known as weak control of the family-wise error rate. As we are using these 40 tests as a method for testing the composite null hypothesis $H_{\uparrow}$ that is true if and only if all of the individual null hypotheses are true, this is sufficient to maintain the test significance.)

There are four possible results of these two hypothesis test, corresponding to all combinations of rejecting or not rejecting $H_{\uparrow}$ and $H_{\downarrow}$. As we already noted, if we reject both hypotheses (and so we reject $H_0$) then we assign ``pass'' to the maximum polarization and ``fail'' to the minimum polarization. Otherwise, we assign the same output for both the maximum and minimum polarization as follows:
\begin{itemize}
\item If we reject $H_{\downarrow}$ but not $H_{\uparrow}$ then we designate both the maximum and minimum polarization as ``pass.''
\item If we reject $H_{\uparrow}$ but not $H_{\downarrow}$ then we designate both the maximum and minimum polarization as ``fail.''
\item If we reject neither $H_{\uparrow}$ or $H_{\downarrow}$ then we designate the maximum and minimum polarization as both ``pass'' (``fail'') if the maximum polarization (minimum polarization) is further from the $
\nicefrac{1}{e}$ threshold than the minimum polarization (maximum polarization). 
\end{itemize}

Alternative techniques for generating frontiers from data may be preferable in other contexts.

\section{Experiment \#2}\label{sec:b2}
The purpose of this appendix is to describe the experiments, and the corresponding data analysis, that are summarized in Figs.~2-3 of the main text. Throughout these appendices we refer to this as \emph{experiment \#2}. This appendix is not intended to be self-contained, and we make explicit references to earlier appendices when necessary. This appendix consists of two parts: in Appendix~\ref{sec:e2-exps} we detail the experiments, and in Appendix~\ref{sec:e2-analysis} we detail the data analysis.

\subsection{Experimental details} \label{sec:e2-exps}
Experiment \#2 used both randomized mirror circuits (see Appendix~\ref{sec:rmcs}) and periodic mirror circuits (see Appendix~\ref{sec:pmcs}) to benchmark each of eight processors and to compare their performance on disordered and ordered circuits. The benchmarking circuits were designed using a procedure that we refer to as \emph{benchmark \#2}. As with benchmark \#1 (see Appendix~\ref{sec:b1}), this procedure can be applied to any gate-model quantum information processor.

\subsubsection{The gate set}\label{sec:b2-gates}
As with benchmark \#1 of experiment \#1, benchmark \#2 is an algorithmic approach for generating mirror circuit benchmarks to run on generic gate-model quantum information processors. It is parameterized by a processor's two-qubit gate $G_2$ and the processor's directed connectivity graph. Unlike benchmark \#1, it uses a standardized single-qubit gate set $\mathbb{G}_1$ consisting of all 24 single-qubit Clifford gates ($\mathbb{C}_1$) for all processors. As in experiment \#1, we used the native two-qubit gate for each processor, which is $\cnot$ for IBM Q processors, and
$\cphase$ for Rigetti processors.

\subsubsection{The circuit shapes}
The first step in the benchmark \#2 algorithm is to select the set of circuit shapes at which to construct benchmarking circuits. For an $n$-qubit processor, we chose the circuit shapes $(w,d)\in\mathbb{W}_n\times\mathbb{D}$, where:
\begin{align*}
&\mathbb{W}_n = \{1, 2, 3, 4, \dots, n\},\\
&\mathbb{D} = \{0, 4, 8, 16, 32, 64, 128, 256, 512\}.
\end{align*}
Exponentially spaced widths would likely be preferable for larger processors, but for the processors we tested, running circuits at an exhaustive set of widths is feasible. For the larger widths we excluded the largest depths, as the processors' error rates implied that all circuits of these shapes would almost certainly all fail. The exact combination of circuit shapes tested can be seen in Fig.~\ref{fig:b2-vbs}. Circuits with depths of 1024 and above were not included only because the IBM Q interface did not allow these circuits to be run.

\subsubsection{The circuit embeddings}
For each processor, we ran width-$w$ circuits on a single set of $w$ qubits. We chose the $w$ qubits predicted to perform the best on our benchmark, according to a simple heuristic based on the processor's published error rates. As is the case for choosing the ``best'' performing qubits from data --- which we did in the data analysis for experiment \#1 (see Appendix~\ref{sec:b1}) --- there are many reasonable ways to use the error rates to choose this qubit set. Using the standard form error rate set $\{\epsilon\}$ introduced in Appendix~\ref{sec:standard-error-rates}, we do so as follows:
\begin{enumerate}
\item We model the success probability for a shape $(w,d)$ benchmarking circuit on the qubit set $q_w$ as
\begin{equation}
S = \left(s(R)- \nicefrac{1}{2^w}\right) \lambda_1^{d(w - \xi)}\lambda_2^{\nicefrac{d \xi}{2}} + \nicefrac{1}{2^w}, \label{eq:q-selection}
\end{equation}
where 
\begin{itemize}
\item $s(R)$ is the success rate of the readout error layer for those qubits, defined in Eq.~\eqref{eq:readout-layer-error},
\item $\xi$ is the target two-qubit gate density of the benchmarking circuits (in these experiments, $\xi = 0$ for $w=0$ and $\xi =\nicefrac{1}{8}$ otherwise),
\item $\lambda_1 =  1 - \frac{4 \epsilon_1}{3}$ where $\epsilon_1$ is the mean error rate of the one-qubit gates on the $s_w$, and
\item $\lambda_2 =  1 - \frac{16 \epsilon_2}{15}$ where $\epsilon_2$ is the mean error rate of the two-qubit gates between the qubits in $q_w$. 
\end{itemize}
This formula is a heuristic for predicting the expected success probability of a shape $(w,d)$ randomized mirror circuit with a two-qubit density of $\xi$.
\item For each connected qubit subset of size $w$, we find the depth $d$ for which Eq.~\eqref{eq:q-selection} predicts that $S=\nicefrac{1}{e}(1 - \nicefrac{1}{2^w}) + \nicefrac{1}{2^w}$. In terms of polarization ($P$) this is the depth at which this equation predicts that $P =\nicefrac{1}{e}$. Note that $d$ is not restricted to being an integer, and it can be negative.
\item For each width $w$, we select the connected qubit subset for which this depth is maximized.
\end{enumerate}

This procedure is one reasonable method for selecting the ``best'' set of qubits using only a set of generic error rates for those qubits --- but note that there are many possible alternative heuristics, and we do not claim our choice is optimal. 

\subsubsection{The circuit sampling}
In this experiment we ran randomized mirror circuits and periodic mirror circuits. As the aim was to investigate the role of circuit order/disorder on circuit failure rates, they were designed to have similar properties. 
\begin{itemize}
\item The randomized mirror circuits were sampled using the edge-grab sampler introduced in Appendix~\ref{sec:e2-sampler}. The \emph{expected} two-qubit gate density was set to $\nicefrac{1}{8}$.
\item The periodic mirror circuits were sampled using the algorithm introduced in Appendix~\ref{sec-germ-e2}. The two-qubit gate density of these circuits is approximately bounded by $\nicefrac{1}{8}$ (it is rigorously bounded by $\nicefrac{1}{8}$ except when a partial repetition of a germ is required --- see Appendix~\ref{sec-germ-e2}).
\end{itemize}
We sampled 40 randomized mirror circuits and 40 periodic mirror circuits for each circuit shape $(w,d)$. As with experiment \#1, the circuits are 
constructed \emph{after} identifying the expected best $w$-qubit set at each width $w$. This allows us to ensure that the benchmark circuits respect connectivity constraints. 

Because the periodic mirror circuits are also randomly sampled from a distribution (see Appendix~\ref{sec-germ-e2}), for the remainder of this appendix we will refer to the randomized mirror circuits in this experiment as \emph{disordered mirror circuits}. The code that we used to perform this sampling has been incorporated into the open-source software package \texttt{pyGSTi} \cite{nielsen2020probing, pygstiversion0.9.9.1}. 

\subsubsection{Experimental details}
We ran benchmark \#2 on eight of the twelve processors that we tested in experiment \#1. The four devices benchmarked in experiment \#1 but not in experiment \#2 (IBM Q Reuschlikon, IBM Q Tenerife, Rigetti Agave and Rigetti Aspen 6) were no longer available once we had designed this benchmark. At the time of these experiments, 8 of the 16 qubits in Rigetti Aspen 4 were not functioning, so it was tested as an 8-qubit processor (whereas all 16 qubits were available when we ran benchmark \#1 on Aspen 4). The experiments were run using the online access services of IBM Q \cite{ibmq2} and Rigetti \cite{rigetti-qcs}. We implemented two ``passes'' through the circuits \cite{rudinger2018probing}: the circuits were looped through with each circuit repeated 1024 times, and then we repeated this a second time (see Fig.~\ref{fig:stability} for more details). Both passes through all the circuits were run within the same calibration window. Unlike in experiment \#1, none of the circuits were run in parallel.

\begin{figure}[t!]
\includegraphics[width=8.4cm]{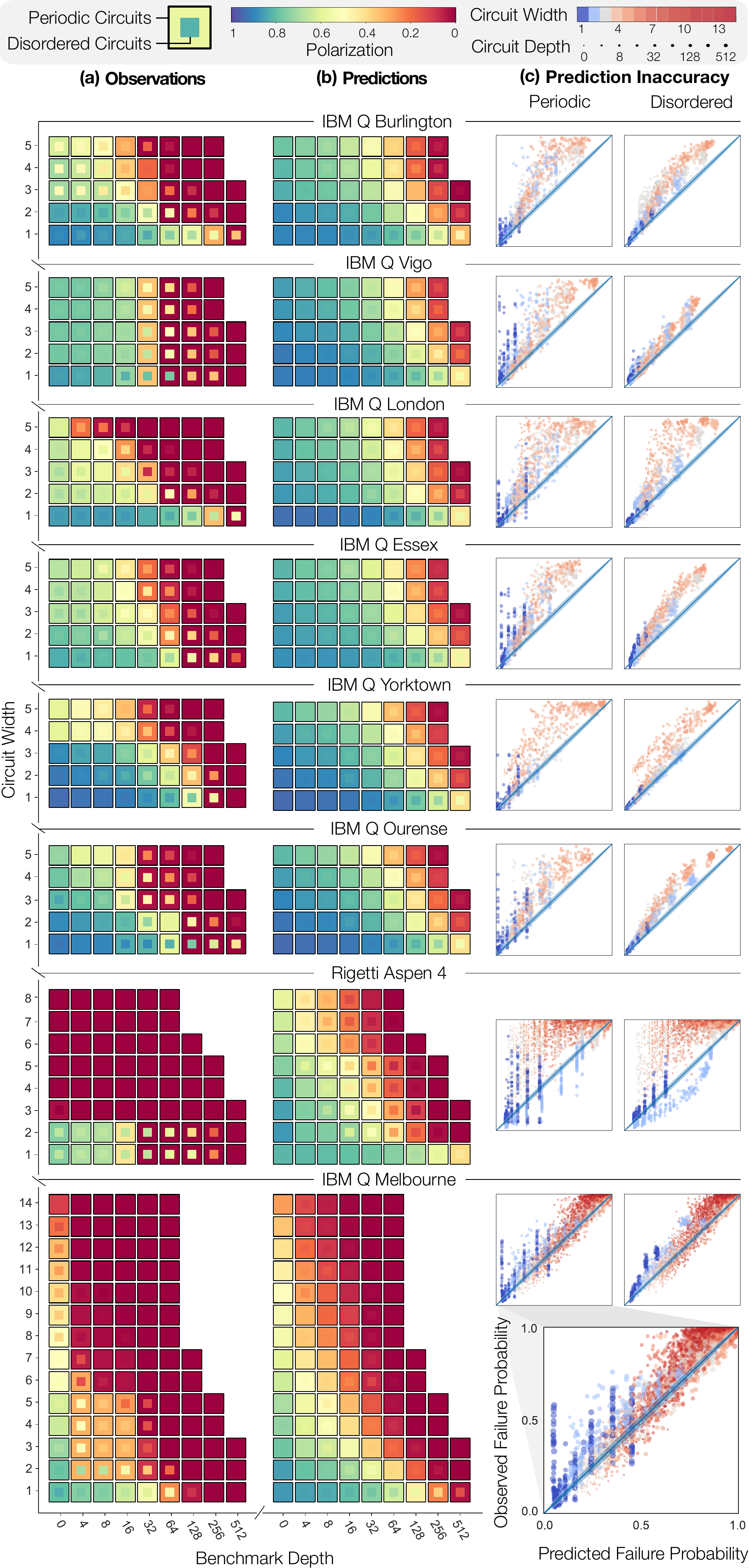}
\caption{{\bf Random benchmarks do not predict structured circuit performance.} This figure includes the results from all of the eight processors benchmarked in experiment \#2, expanding on Fig.~2 from the main text (which shows only the plots in the lower four rows).~{\bf (a)} The output polarization versus circuit width and depth for periodic (outer squares) and disordered (inner squares) circuits, minimized over all the test circuits that have that width or less and that depth or less.~{\bf (b)} Predictions from each device's error rates, accounting for the finite repeats of each circuit in the experiment  ($N=1024$) via a standard bootstrap.~{\bf (c)} The observed versus predicted failure rates for every circuit that was run. The blue diagonal bands are $2\sigma$ confidence regions: if the predictions were correct, $\approx95\%$ of the data would fall in them.}
\label{fig:b2-vbs}
\end{figure}

\begin{figure*}[t!]
\includegraphics[width=18cm]{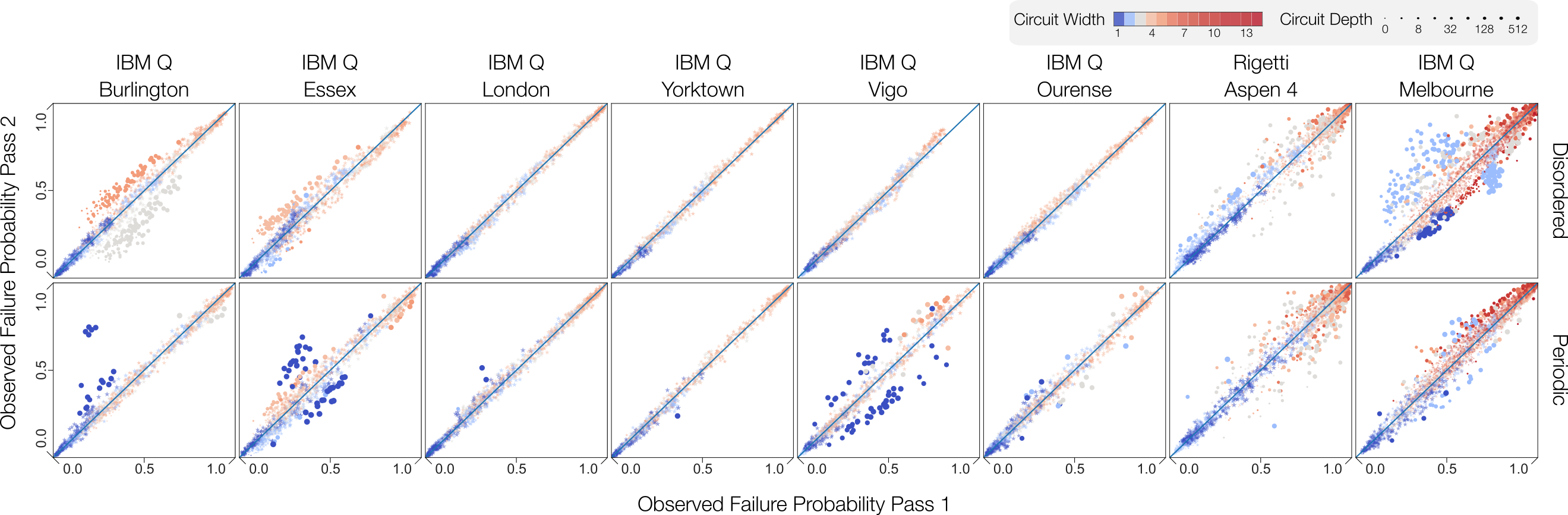}
\caption{\textbf{Quantifying temporal instability}. Experiment \#2 consisted of both periodic and disordered mirror circuits. These circuits were run in four batches in the following order: all the periodic circuits (pass 1), all the disordered circuits (pass 1), all the periodic circuits (pass 2), all the disordered circuits (pass 2). We thus obtained observed success probabilities, $\hat{S}_1$ and $\hat{S}_2$, for each circuit in passes 1 and 2, respectively. For each processor, this figure plots each circuit's observed failure rate in the first pass $(1-\hat{S}_1)$ versus the circuit's observed failure rate in the second pass $(1-\hat{S}_2)$. The upper (lower) row shows data from the disordered (periodic) circuits. Solid circles (translucent stars) are data for which the difference between the failure rates in the two passes is (is not) 5\% statistically significant as assessed using the technique of Ref.~\cite{rudinger2018probing}. For further details, see Appendix~\ref{sec:stability}.} 
\label{fig:stability}
\end{figure*}

\subsection{Data analysis}\label{sec:e2-analysis}
The results of experiment \#2 are summarized in Figs.~2 and 3 of the main text. For brevity, Fig.~2 shows the results for only four of the eight benchmarked processors, so in Fig.~\ref{fig:b2-vbs} we expand on Fig.~2 to show the results for all eight processors. In the remainder of this appendix we explain the data analysis used to generate these plots. We use notation that explicitly distinguishes between a circuit's true success probability ($S$) and the observed success probability ($\hat{S}$) obtained from a finite number of repetitions of that circuit. Because the data was taken in two passes, it is useful to further distinguish between a circuit's true and observed success probability at the time of the first pass through the circuits ($S_1$ and $\hat{S}_1$, respectively) and a circuit's true and observed success probability at the time of the second pass through the circuits ($S_2$ and $\hat{S}_2$, respectively). If the processor is stable then $S_1 = S_2$ for every circuit, but this is not guaranteed to be true, as drift is a common problem in quantum processors \cite{rudinger2018probing, proctor2019detecting, wan2019quantum}.

\subsubsection{Quantifying processor instability}\label{sec:stability}
The first step in our data analysis identifies instability in the processors by comparing  the two passes through the benchmarking circuits using the statistically rigorous hypothesis testing technique of Ref.~\cite{rudinger2018probing}. Although the results of this analysis are an informative performance benchmark in their own right, this analysis was primarily implemented so that the presence (or absence) of detectable instability could be used to inform other aspects of the data analysis. The formal aim of this analysis is to assess whether there is statistically significant evidence in the data that $S_1 \neq S_2$ for any circuit. The analysis performs statistical hypothesis tests of the null hypothesis that $S_1 = S_2$ for each circuit. Fig.~\ref{fig:stability} plots $(1-\hat{S}_1)$ versus $(1-\hat{S}_2)$ for every periodic (upper row) and disordered (lower row) mirror circuit that was run on each processor (the columns). Solid circles (transparent stars) denote the observed failure probabilities that are (are not) sufficiently different to constitute statistically significant evidence that $S_1 \neq S_2$ for that circuit, according to the hypothesis tests of Ref.~\cite{rudinger2018probing}. (The procedure of Ref.~\cite{rudinger2018probing} is designed for strong control of the family-wise error rate. We implemented the hypothesis test at $5\%$ significance. The test significance is not corrected to account for the fact that we are testing eight different processors.)

There is statistically significant evidence in the data that all eight processors are unstable between the two passes through the circuits, \ie, for every processor there is evidence that there is at least one circuit for which $S_1 \neq S_2$. However, the magnitude of the instability varies dramatically between processors. For example, the difference between $\hat{S}_1$ and $\hat{S}_2$ is small for \emph{every} circuit that was run on IBM Q Yorktown, whereas many circuits exhibit large differences between $\hat{S}_1$ and $\hat{S}_2$ for IBM Q Melbourne and Rigetti Aspen 4. We note that the experiment on IBM Q Melbourne took approximately 10 hours whereas all other experiments took under 3.5 hours, so the difference between the observed instability on IBM Q Melbourne and the other IBM Q devices should not be used to infer that IBM Q Melbourne suffered from worse instabilities than the other IBM Q devices. As we explain further below, because of the results of this instability analysis we discarded the data from the second pass through the circuits --- \ie, we used only the data from the first pass in the remainder of the analysis.

\subsubsection{Worst-case volumetric benchmarks}
In Fig.~\ref{fig:b2-vbs}a (and Fig.~2a) we summarized the difference between the success rates of periodic and disordered circuits in terms of worst-case performance. For each processor and each benchmarked shape $(w,d)$, Fig.~\ref{fig:b2-vbs}a (and Fig.~2a) shows the observed polarization versus circuit shape $(w,d)$ for periodic (outer squares) and disordered (inner squares) circuits, minimized over all the test circuits of shape $(w^*,d^*)$ where $w^* \leq w$ and $d^* \leq d$. This was calculated using only the data from the first pass through the circuits.

The observed minimum polarization is a biased estimate for the true minimum polarization over a circuit ensemble. If each circuit's success probability was stable over time, this bias could be removed by using the data from the first pass through the circuits to select the worst-performing circuit for each circuit shape, and then using the data from the second pass to estimate these circuits' polarizations. However, the validity of that strategy is based on the assumption of stability, and there are large instabilities between the two passes for some processors (see Fig.~\ref{fig:stability}). As the purpose of Fig.~2a is to \emph{compare} periodic and disordered circuits, and we ran the same number of periodic and mirror circuits of each shape and we ran every circuit the same number of times, we therefore chose to make no adjustment for finite sampling in the analysis for Fig.~2a.

\subsubsection{Comparing to the predictions of each processor's error rates}
In Fig.~\ref{fig:b2-vbs}b-c (and Fig.~2b-c) we compare our experimental results to predictions derived from each processor's published error rates. The method used to predict the success probability for a specific circuit is explained in Appendix~\ref{sec:predictions}. Fig.~\ref{fig:b2-vbs}c simply plots the predicted failure probability (\ie, one minus the predicted success probability) against the observed failure probability $(1 - \hat{S})$ for every circuit that we ran. This is arranged by processor (the rows) and is further split into periodic and disordered mirror circuits (the left and right columns, respectively). As with all of Fig.~\ref{fig:b2-vbs}c, we include only the data from the first pass through the circuits.

In Fig.~\ref{fig:b2-vbs}b we show volumetric benchmarking plots of the \emph{predicted} worst-case performance implied by the processors' error rates. As discussed above, the analysis resulting in Fig.~\ref{fig:b2-vbs}a does not correct for finite sampling bias in the estimate of each of the minimum polarizations. To ensure that Fig.~\ref{fig:b2-vbs}b may be fairly compared to Fig.~\ref{fig:b2-vbs}a, we simulate this bias using a standard parametric bootstrap. For each processor:
\begin{enumerate}
\item We generated 1000 bootstrapped data sets, by sampling an ``observed'' success probability for each circuit that we ran, given by the average of 1024 draws from a 0/1 valued ``coin'' with the success probability set to the predicted success probability of the circuit.
\item For each bootstrapped data set, we implemented exactly the same analysis that was applied to the experimental data to generate Fig.~\ref{fig:b2-vbs}a. This analysis computes a statistic $\lambda(w,d)$ at each circuit shape $(w,d)$. This results in 1000 bootstrapped values for each $\lambda(w,d)$. 
\item The predicted $\lambda(w,d)$ is then set to the mean of the 1000 bootstrapped values.
\end{enumerate}
\subsubsection{Empirical capability regions}
In Fig.~3 of the main text, we summarize the performance of all eight processors on both periodic and disordered mirror circuits, by dividing the circuit width $\times$ depth plane into ``success'', ``indeterminate'', and ``fail'' regions. These regions correspond to the circuit shapes at which all, some, and none of the 80 test circuits succeeded, respectively, where a circuit is considered to succeed if $P \geq \nicefrac{1}{e}$. To estimate these regions from the data we use statistically hypothesis testing. We start from the null hypothesis that, at shape $(w,d)$, every circuit succeeds ($P \geq \nicefrac{1}{e}$) or every circuit fails ($P < \nicefrac{1}{e}$), and we assign a circuit shape to ``indeterminate'' only if statistical hypothesis testing on the data allows us to reject this null hypothesis with 5\% statistical significance. 

The statistical hypothesis testing is performed using the same technique that we used to generate the performance frontiers from the data in experiment \#1 (this is described in detail in Appendix~\ref{sec:hypothesis}), and the analysis uses only the data from the first pass through the circuits. The reason for using this hypothesis testing framework is that it addresses a bias towards including \emph{every} circuit shape in the ``indeterminate'' region. To understand this, consider running $K$ circuits of shape $(w,d)$ with each circuit repeated $N$ times and with these $K$ circuits sampled from some circuit ensemble in which every circuit has a success probability that is not exactly 1 or 0. Then, for fixed $N$, the probability of the \emph{observed} polarization being above $\nicefrac{1}{e}$ for at least one of these $K$ sampled shape $(w,d)$ circuits and being below $\nicefrac{1}{e}$ for at least one of these $K$ sampled shape $(w,d)$ circuits converges to 1 as $K$ increases, even if every circuit in the ensemble has a success probability well above the threshold value. 
\end{document}